\documentclass[letterpaper,twocolumn,10pt]{article}
\usepackage{usenix-2020-09}
\usepackage{tikz}

\usepackage{filecontents}

\usepackage{microtype}
\usepackage{graphicx}
\usepackage{caption}
\usepackage{subcaption}
\usepackage{booktabs} 

\usepackage{amssymb}
\usepackage{listings}
\usepackage{array}
\usepackage{xspace} 
\usepackage[pdf]{graphviz}
\usepackage[normalem]{ulem} 
\usepackage{comment}
\graphicspath{ {./figures/} }
\usepackage{varwidth}
\usepackage[ruled, linesnumbered]{algorithm2e}
\usepackage{enumitem}
\usepackage{footnote}
\usepackage{multirow}
\usepackage{tablefootnote}
\usepackage{xpatch}
\usepackage{algorithm2e}
\usepackage{balance}
\usepackage{makecell}
\usepackage{amsmath}
\usepackage{amsthm}
\theoremstyle{definition}  
\newtheorem{example}{Example}  

\usepackage[normalem]{ulem}

\usepackage{listings, xcolor}
\usepackage{hyperref}

\date{}

\begin{document}
\title{\Large \bf EdgeServe: A Streaming System for Decentralized Model Serving}
\newcommand{\sys}{EdgeServe\xspace}



 \author{
 {\rm Ted Shaowang}\\
 The University of Chicago
 \and
 {\rm Sanjay Krishnan}\\
 The University of Chicago
 }
\maketitle

\begin{abstract}
The relevant features for a machine learning task may arrive as one or more continuous streams of data. Serving machine learning models over streams of data creates a number of interesting systems challenges in managing data routing, time-synchronization, and rate control. This paper presents \sys, a distributed streaming system that can serve predictions from machine learning models in real time.
We evaluate \sys on three streaming prediction tasks: (1) human activity recognition, (2) autonomous driving, and (3) network intrusion detection.
\end{abstract}

\newcommand{\todo}[1]{\textcolor{red}{\bf [TODO: #1]}}



\section{Introduction}
The broader computing community has long understood the importance of telemetry in both physical and digital systems.
The growing maturity of AI has created new opportunities for such data, where models can be built to predict future behavior and/or automatically react to current trends -- to ``close the loop''.
This paper presents \sys, a new system that allows for low-latency feedback systems over distributed streams of data. 

Efficiently serving predictions from machine learning models is already a crucial part of modern software applications ranging from automatic fraud detection to predictive medicine~\cite{tfcasestudies}. 
Accordingly, a number of \emph{model serving frameworks} have been developed, including Clipper~\cite{clipper}, TensorFlow Serving~\cite{tfserving}, and InferLine~\cite{inferline}. 
These frameworks simplify the deployment and interfacing of trained machine-learning models with a service-oriented interface.
Typically, they provide a RESTful API that accepts features as inputs (i.e., a prediction ``request''), and responds to these requests with predicted labels (i.e., a prediction ``response'').
These frameworks provide a number of crucial optimizations such as containerizing inference code~\cite{clipper}, autoscaling~\cite{tfserving}, and model ensembling~\cite{inferline}.

Existing model serving frameworks were envisioned as components in cloud-based deployments.
Implicit to this design, there are several key assumptions: (1) prediction requests arrive asynchronously through the RESTful interface, (2) the request is self-contained with all of the features necessary to issue a prediction, (3) the design prioritizes scalability over the latency of an individual request, (4) and the response is delivered back to the requester.
We find that streaming settings challenge this design paradigm.
Consider a simple example of a model where the time-ordering of predictions matters (e.g., sensor fusion or forecasting).
If such a model is served with a RESTful model-serving framework, there is no inherent message ordering guarantee which is crucial for accurate forecasting.
The data processor needs to block processing until a prediction is returned by the framework, and this negates any pipelining or scale-out optimizations present in these frameworks. 
For such use cases, it is more convenient for developers to think of a machine learning model as an operator applied to one or more continuous streams of data with synchronization, rate limit, and freshness constraints.

To the best of our knowledge, the academic literature on this topic is relatively sparse with most existing work in video analytics~\cite{zhao2023streaming, kang2017noscope, arulraj2022accelerating, horchidan2022evaluating,flinkml, tfkafka}. 
There is also a significant amount of work in real-time systems~\cite{flink, spark-structured-streaming, apache-samza, apachestorm}, but few systems focus on model serving.
In particular, significant technical challenges arise when the relevant features for a machine learning model are generated on different network nodes than where the model is served.
The data has to get to ``the right place at the right time'' before any prediction can be made, and this communication quickly becomes the primary bottleneck.
The problem is further complicated where there are multiple data streams: the data streams have to be time-synchronized and integrated before any predictions can take place.
Prior work has shown that placement and synchronization decisions affect both performance and accuracy in nuanced ways~\cite{shaowang2021declarative, shaowang2022bidede}.
Thus, for low-latency model-serving over distributed streams of data, one has to jointly optimize for communication, rate control, and the fact that minor time alignment deviations typically have a negligible impact on accuracy in model-serving scenarios.

To better understand the tradeoffs, consider the following running example.

\begin{example}
In network intrusion detection, machine learning models applied to packet capture data are used to infer anomalous or malicious traffic patterns. Most organizations have geo-distributed private networks spanning multiple clouds and regions. The relevant features for a particular intrusion detection model may be sourced from different packet capture streams at different points in the network.
These streams will have to be synchronized and integrated to make any global prediction.
\end{example}

With existing tools, building such applications requires significant developer effort in the design of (1) communication between nodes collecting the streams, (2) the time-alignment strategy for the streams, and (3) the rate control of incoming data.  Challenge (1,2,3) create a complex tradeoff space that leads to bespoke solutions~\cite{IntelPress2020, tfcasestudies, tfkafka, lyft}.
This paper describes a first step towards such a system, called \sys, that addresses this need.
Instead of a RESTful service that handles each prediction request asynchronously, \sys routes synchronized streams of data to models that are flexibly placed anywhere in a network.
We call such an architecture \emph{synchronized prediction} to differentiate it from classical model serving, where a collection of model-serving nodes work together to serve predictions over one or more data streams in a temporally coherent way.

Practically, \sys provides a lightweight inference service that can be installed on every node of the network.
\sys employs a message broker to route data around different nodes, allowing multiple producers and consumers to operate on the same message queue simultaneously. Users can define data movements and model placements by pointing models to named streams of data rather than their physical locations. Furthermore, the user can program her model and featurization as if there was all-to-all communication in the network, and the actual data routing over the actual network topology is handled seamlessly by \sys.
These data streams can be time-synchronized so that inferences that need to look at a particular snapshot in time can appropriately construct features that join data from different sources.
Furthermore, the data can be derived from primary sources (e.g., sensors, user data streams, etc.) or can be results of computation (e.g., features/predictions computed from pre-trained models).
This flexibility allows users to build complex but robust predictive applications in networks with heterogeneous and disaggregated resources. 

While \sys resembles other streaming and data flow systems~\cite{goog_data,spark,spark-structured-streaming,naiad,apache-samza,apachestorm,twitter-heron,TensorFlow,pytorch,flink}, there are three key novel architectural features due to the model-serving focus.
\begin{itemize}
    \item (How to trigger computations?) \emph{Data-Triggered Stream Joins.} \sys employs a novel temporal join strategy for combining multiple streams of data based on data arrivals (\S\ref{sec:join}). 

    \item (What are the communication primitives?) \emph{Lazy Data Routing.} For large data payloads (e.g., high-dimensional data streams), \sys applies an innovative communication protocol called ``lazy data routing'' where only references to data are sent through the message broker. (\S\ref{sec:lazy})

    \item (How to ensure reliable behavior?) \emph{Prediction Rate Control.} \sys presents strategies that can ensure that timely decisions still get made even in the presence of dropped, delayed messages or overloaded models. (\S\ref{sec:rate-control})
\end{itemize}

\section{Background and Existing Model Serving Frameworks}
This section motivates \sys and describes the performance of existing model serving frameworks.

\begin{example}
To understand how existing cloud-based systems work, we construct a simplified scenario where a single stream of data is fed into a model-serving framework. Each data item is a 134-dimensional feature vector, the model-serving framework must issue a prediction for each item. The items are streamed into a message broker and dequeued in timestamp order. The goal of this experiment is to illustrate that queueing and communication far outweigh the actual model inference time for typical sensing workloads.
\end{example}

Based on blog posts and tutorials that describe best practices, we developed a few different models serving pipelines on AWS and GCE \cite{gce}. We experimented with two different models, a Random Forest and a 3-layer MLP. We used roughly comparable inference hardware on both cloud providers (on AWS SageMaker EC2 P3 and GCE a VertexAI 2.10 Container), and note that this inference hardware is GPU-accelerated.

\begin{itemize}
    \item \textbf{AWS. } This model-serving pipeline uses AWS SQS to queue messages and AWS SageMaker to perform the inferences over each queued message. 
    \item \textbf{GCE. } This model-serving pipeline uses GCE Pub/Sub to queue messages and GCE VertexAI to perform the inferences over each queued message.
    \item \textbf{Inf Only. } We run both the AWS and GCE pipelines above in an inference-only mode which only measures the latency of AWS SageMaker and GCE VertexAI repsectively.
\end{itemize}

We evaluate these two baselines in terms of their end-to-end latency, which is the elapsed time since the execution of the ``publish'' message to the message queue and the delivered prediction. 
\begin{table}[ht!]
\begin{tabular}{|l| l|l||l|l|}
\hline
    & \multicolumn{2}{c||}{Random Forest}          & \multicolumn{2}{c|}{MLP}                                                                \\
    & Med. &P99 & Med. &P99 \\ \hline\hline
AWS & 141ms & 400ms & 116ms & 391ms \\
AWS (inf only) & 20ms & 25ms & 18ms & 20ms \\
GCE & 88ms & 94ms    & 74ms & 82ms \\ 
GCE (inf only) & 18ms & 22ms    & 13ms & 15ms \\  \hline
\end{tabular}
\caption{End-to-end latency for model inference over a 134-dimension sensor stream}
\end{table}

In terms of end-to-end latency, existing frameworks are not satisfactory for emerging ``real-time'' machine learning applications, where the typical latency tolerance is measured in tens of milliseconds.
With default cloud tools, one can expect hundreds of milliseconds of latency. Even worse, this is often highly unpredictable.
Interestingly enough, the primary source of end-to-end latency is not the model inference itself, but delays in message queuing. 
Streaming the data to the model becomes a bottleneck, incurring copying costs, queuing costs, and checkpointing/replication costs at the message broker.
Cloud-based messaging services were designed to be highly available and reliable, but not particularly aimed for low-latency or time-synchronized applications.
These queuing overheads can be made arbitrarily more significant if multiple streams of data are required. 
Then, there is additional waiting time in the system to align observations across streams.

\subsection{What is Streaming Inference?} \label{sec:streaming-inference}
These numbers indicate the need for a new model serving framework that tightly \textbf{integrates streaming with model inference.} 
We can build a serving framework that is more suited for streaming data, and we can build a streaming system that is more suited for the typical workloads seen in machine learning serving.

\noindent \textbf{Inference over a Single Stream. } Consider a supervised learning inference task. Let $x$ be a feature vector in $\mathcal{R}^p$ and $f_\theta$ be a model with parameters $\theta$. $f_\theta$ evaluates at $x$ and returns a corresponding prediction $y$, which is in the set of labels $\mathcal{Y}$. A prediction over a stream of such feature vectors can be thus summarized as:
\[ y_t = f_\theta(x_t) \]
where $t$ denotes a timestamp for the feature vector. In such a prediction problem, the user must ensure that the featurized data is at ``the right place at the right time'': $f_\theta$ has to be hosted somewhere in a network and $x_t$ has to be appropriately generated and sent to $f_\theta$.

\noindent \textbf{Inference over Multiple Streams. } Now, let's imagine that $x_t$ is constructed from multiple different streams of data. Each $x_t$ (the original features) can be treated as a concatenation of $d$ individual streams:
\[
x_t = \begin{bmatrix}
x^{(1)}_t &
... &
x^{(i)}_t &
... &
x^{(d)}_t  
\end{bmatrix} \
\]
Each of these streams of data $x^{(i)}_{1},...,x^{(i)}_t$ might be produced on a different node in a network.
Consider the network intrusion detection example (Example 1). Each $x^{(i)}$ corresponds to one of the streams of data (packets from node 1, packets from node 2, packets from node 3).
In this case, we have different streams of data $x^{(1)},x^{(2)},...$ coming in, and we need to aggregate them so that the final prediction arrives in our desired destination node.

If the streams of sub-features are collected independently, they will likely not be time-synchronized. This means, at any given instant, the data at the prediction node comes from a slightly different timestamp:
\[
x_t = \begin{bmatrix}
x^{(1)}_{t+\epsilon_1} &
... &
x^{(1)}_{t+\epsilon_i} &
... &
x^{(d)}_{t+\epsilon_d}  
\end{bmatrix} \
\]
Each $\epsilon_i$ denotes a positive or negative offset. The overall time-skew of the prediction problem is $\epsilon = \text{max}_{i} \epsilon_i - \text{min}_{j} \epsilon_j$. In other words, to issue a perfectly synchronized prediction at time $t$, the earliest stream has to wait for $\epsilon$ steps to ensure all features are available.
This can be even more complicated if different data streams are collected at different frequencies.
\sys provides an API for controlling synchronization errors in decentralized prediction deployments (\S\ref{sec:rate-control}).

\noindent \textbf{\sys: Our Contribution. }
Today's model serving systems lack the support for flexibly deploying models (or partial models) across a network and routing data and predictions to/from them.
Users with such problems today have to design bespoke solutions, which can result in brittle design decisions that are not robust to changes in the network or data.
While it is true that prior work has considered decomposing models across a network to optimize throughput~\cite{narayanan2019pipedream}, this work does not consider latency-sensitive applications nor does it consider disaggregated input data streams. 
\sys significantly reduces latency in queuing and communication leading to large improvements in end-to-end responsiveness. 
To motivate the contributions, if we run the same experiment on comparable hardware in AWS as above in \sys, we get the following results:

\begin{table}[ht!]
\begin{tabular}{|l| l|l||l|l|}
\hline
    & \multicolumn{2}{c||}{Random Forest}          & \multicolumn{2}{c|}{MLP}                                                                \\
    & Med. &P99 & Med. &P99 \\ \hline\hline
\sys & 21ms & 31ms & 20ms & 29ms \\ \hline        
\end{tabular}
\caption{End-to-end latency For model inference over a 134-dimension sensor stream using \sys}
\end{table}

\section{\sys Architecture and API}
\label{engineering}
\sys is a system that facilitates prediction applications on streaming data.

\subsection{\sys Overall Workflow}
The key difference between \sys and model serving systems is that \sys takes streams as the unit of operations.
Existing model serving services all follow the request-response API design, which requires that the prediction results be sent back to the caller.
This might not be suitable for streaming use cases, as users likely prefer results elsewhere for decision-making purposes.
\sys simply routes results to another message topic that could be consumed by any node.

Existing model serving systems use RESTful APIs to handle inputs and outputs: they take each input data item as an HTTP request and issue a prediction as the HTTP response back to the caller.
Requests and responses always appear in pairs. This 1:1 relationship makes it impossible to issue one multi-modal prediction based on multiple data inputs.
The only workaround would be to join those data manually and send the joined result as the HTTP request.
\sys, on the other hand, uses message queues to route data around. Each data modality forms its own message queue and they can be joined in real time as tuples before passing to the model. The prediction result, depending on how many outputs it consists of, is routed to one or more message queues for downstream operators to consume.

\subsection{Execution Layer API}
\begin{figure}[t]
    \centering
    \includegraphics[width=0.8\columnwidth]{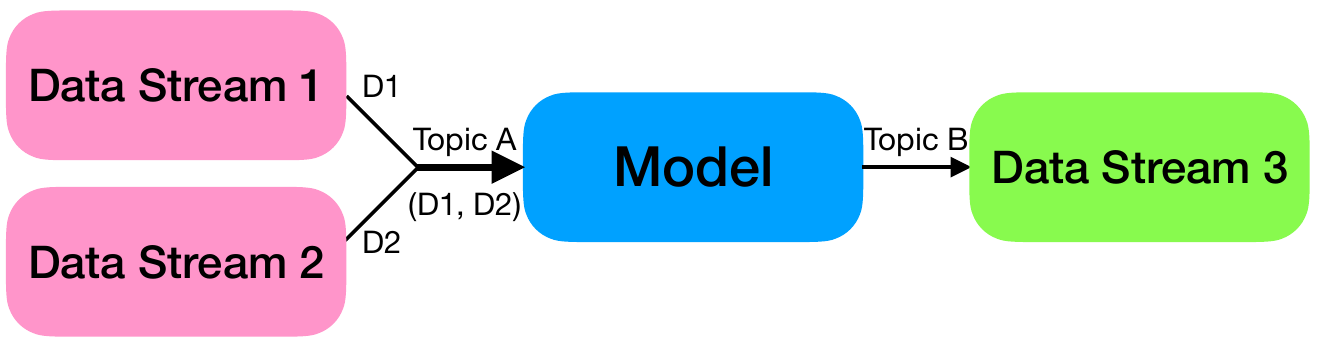}
    \caption{\sys's execution layer. Data from data streams 1 and 2 (D1, D2) are paired together since both streams are grouped into topic A.}
    \label{fig:data-stream-model}
\end{figure}
\sys runs as a process on every node in the network. 
Every node running \sys can potentially create and consume data streams, and run model inference.
One of the nodes is designated as the \emph{leader node} running our message broker backend.
\sys extends Apache Pulsar~\cite{apachepulsar} to build a low-latency message broker backend to transfer messages between nodes.
This is the node that coordinates message routing and maintains a canonical clock for the network.
This leader can be selected through a leader election algorithm (e.g., ~\cite{malpani2000leader}), or can simply be selected by the user.
The leader is also responsible for dispatching user-written code to the other nodes on the network.
\sys assumes that these nodes are connected via a standard TCP/IP network and every node can directly communicate with the leader.
While \sys does not require all-to-all communication, having this capability can be advantageous, especially with large payloads.
This is because an optimization technique known as \textit{lazy data routing} (\S\ref{sec:lazy}) utilizes all-to-all communication.

\vspace{0.5em} \noindent \textbf{Data Streams API: }
Any node on the network, including the leader, can register globally-visible data streams to the network.
All data in \sys are represented as infinite streams of data. 
These streams can be of any serializable data type and leverage Python iterator syntax.
To invoke \sys, the user simply needs to wrap each data stream as a Python generator and register the stream with the leader.
Other nodes on the network can read from this stream of data by accessing an iterator-like interface.

Streams are further grouped into ``topics'' representing joint predictive tasks, as illustrated in Figure~\ref{fig:data-stream-model}.
For example, the streams from ``packet capture 1'' and ``packet capture 2'' could be combined for a particular model.
Grouping streams into topics gives the system information on which streams have to be synchronized and joined together.
Each message contains details about its originating data stream and associated topic.

\vspace{0.5em} \noindent \textbf{Models API: }
Over these streams of data, we would like to compute different machine learning inferences. A ``model'' object encapsulates such computation. A model consumes one or more input data streams from the same topic, and outputs one or more data streams.
We take a general view of what a model is: a model is simply a unit of computation that is performed synchronously over a stream of data.
In \sys, a ``model'' is just an operator that produces predictions triggered by the input streams. 
This stream of predictions can be further combined into topics that other models consume. 
The same model object can represent a sub-model (e.g., one member of an ensemble), or a featurizer (e.g., a function that computes a set of features).

Models that process these streams take multiple data streams as input, e.g., a multi-modal model. These data streams need to be temporally synchronized before going to the model.
This logic is encapsulated in a component called a ``joiner''.
A joiner fills the gap between data streams and the multimodal model.
It consumes data from multiple streams and produces a single iterator interface for models.
We discuss more details on the joiner in \S\ref{sec:join}.

Our model API is specifically designed to simplify decentralized deployments where the output of one set of models is consumed by others (e.g., an ensemble). 
We treat ensembling just like another model, which takes other models' predictions as inputs, and our system is able to combine them together in a time-synchronized way. 
Users only need to focus on the actual ensembling algorithm and leave communication and placement details to our system.

\section{Highly-Responsive Data Stream Joins}\label{sec:join}
In typical time-series databases, band-joins are used to integrate such series~\cite{dewitt1991evaluation, khayyat2015lightning}, where all items within a certain time-delta are grouped together.
True band-joins are challenging in streaming systems where data may arrive out-of-order or in a bursty way leading to potentially unbounded buffering, so existing streaming systems offer an approximation using tumbling time windows~\cite{flink, apachestorm, apache-samza, spark-structured-streaming,photon,facebook-streaming-join}.
All items that fall within the same tumbling time window are grouped together.
We refer to this type of streaming join as a \textit{time-triggered join}, i.e, the join condition is triggered by a clock tick.

This type of join can cause delays that affect the timeliness of predictions.
A multimodal model's response time to new data is limited by the width of the tumbling window.
An alternative to time-triggered joins is to join eagerly as soon as a new item is published to the stream, which we call a \textit{data-triggered join}.
The novelty of this approach is that it is responsive to new data, while having a bounded buffer size.
We will discuss both join methods in more detail below.

\subsection{Time-triggered Join}\label{sec:time-join}
As the name suggests, a time-triggered join buffers incoming messages from all streams over a time window, and triggers a join result at the end of the time window.
Within each time window, only the latest message from each stream is kept as that stream's input. The definition of `latest' here can be either event time or processing time. These latest messages are combined as a tuple before sending downstream.

\begin{figure}[t]
    \centering
    \begin{minipage}{.5\columnwidth}
      \centering
      \includegraphics[width=\linewidth]{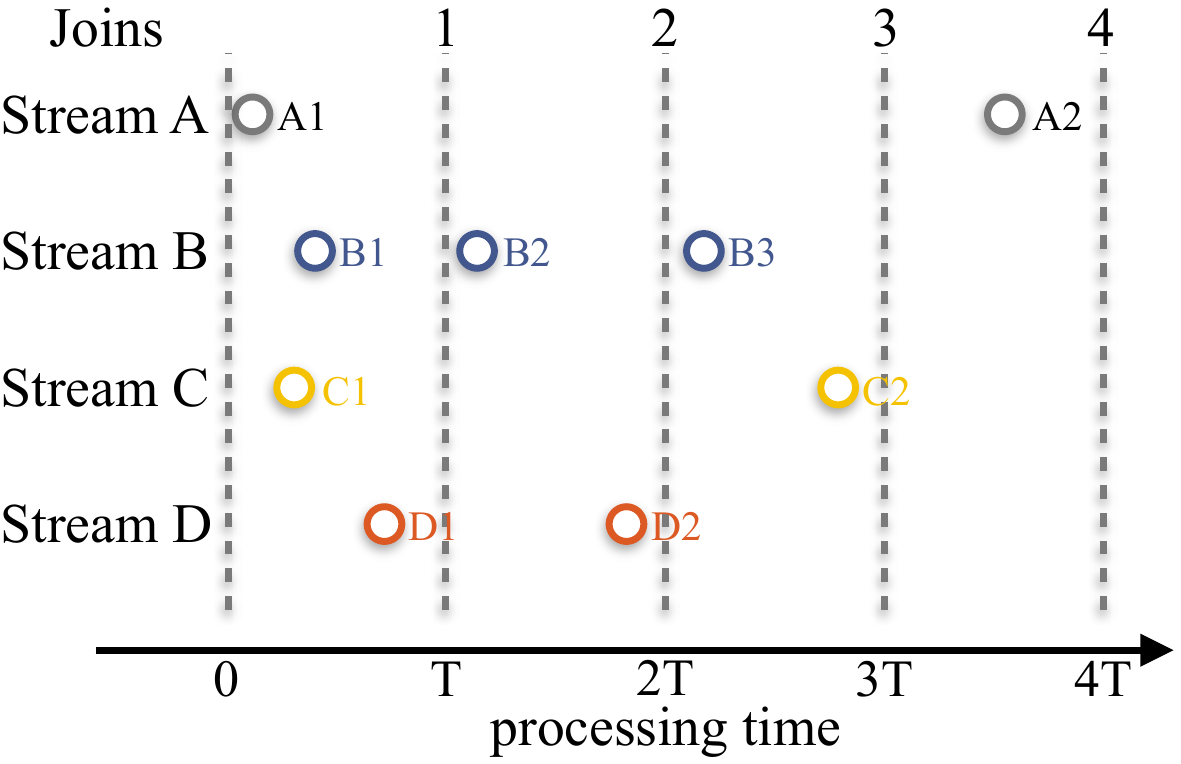}
      \caption{Time-triggered join.}
      \label{fig:time-triggered-join}
    \end{minipage}%
    \begin{minipage}{.5\columnwidth}
      \centering
      \includegraphics[width=\linewidth]{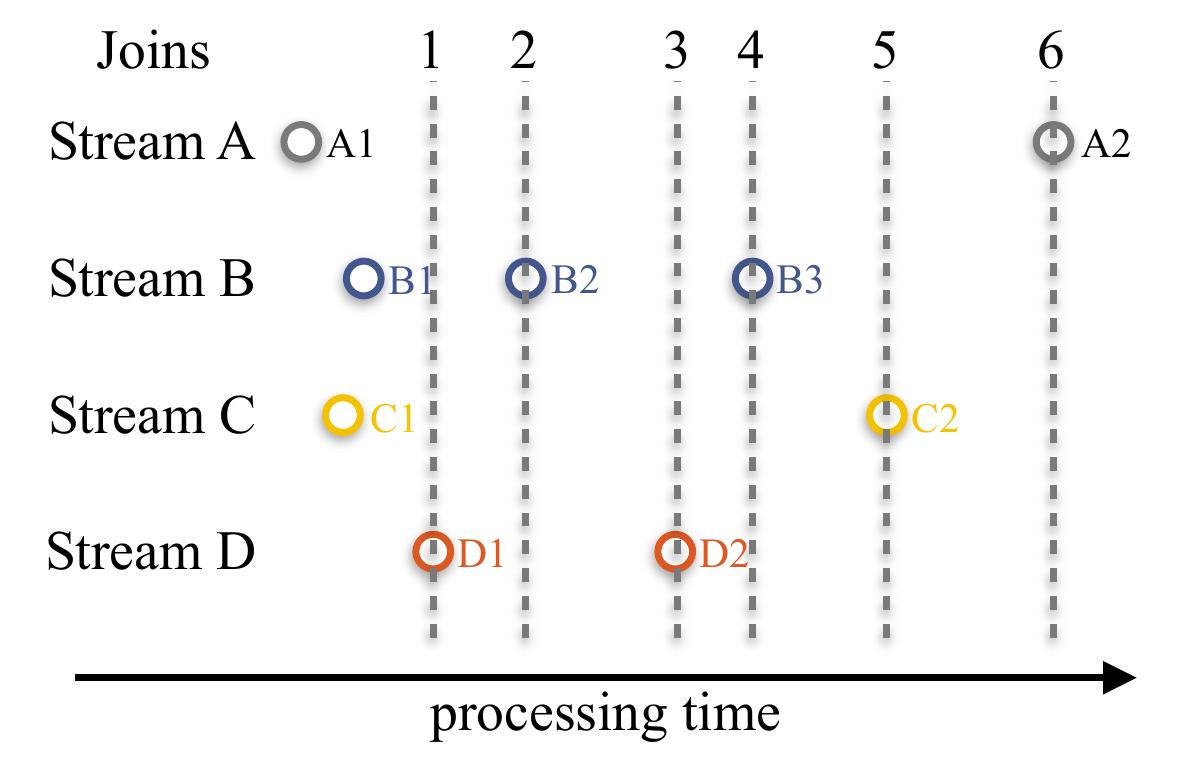}
      \caption{Data-triggered join.}
      \label{fig:data-triggered-join}
    \end{minipage}
\end{figure}
Figure~\ref{fig:time-triggered-join} is an example of a time-triggered join.
In this example, we assume event time and processing time are the same for simplicity. The join results are as follows:
join 1 (A1, B1, C1, D1);
join 2 (A1, B2, C1, D2);
join 3 (A1, B3, C2, D2);
join 4 (A2, B3, C2, D2).
Intuitively, waiting for a time-triggered join resembles waiting for a bus.
Since B2 arrives immediately after the join at $t=T$ was issued, it will have to wait until $t=2T$ to get processed, resulting in a longer waiting time.
On the other hand, time-triggered joins are beneficial when joins are desired at fixed frequencies, as they smooth out the burstiness of incoming data.

The state management for a time-triggered join is rather straightforward.
If the join is based on processing time, only the latest messages from each stream need to be buffered.
If the join is based on event time, all messages within a fixed time window need to be additionally buffered, in case they arrive out of order in terms of event time.

\subsection{Data-triggered Join}\label{sec:data-join}
An alternative to a time-triggered join is to perform a join whenever a new piece of data from any stream arrives.
Intuitively, the latest known data from all streams are buffered in order to join with the \underline{new data item} (underlined in the following example). We will show how this works precisely in the two-way case, and it should be clear how to extend this to a multi-way join.

Given two streams StreamA and StreamB, the algorithm tracks the latest known item from each stream and its timestamp.
Each time the stream publishes a new data item, it is joined with the latest known item from the other stream.
As before, the timestamp can refer to either event time or processing time.
However, the order of joined tuples is not guaranteed in terms of event time, since the joining process depends on when the data is actually received by the joiner.

\vspace{0.25em}
\noindent\fbox{%
\footnotesize
    \parbox{0.8\columnwidth}{%
        \noindent \textbf{Data-Triggered Join Algorithm} \\

\textbf{Given: } StreamA, StreamB 

\textbf{Set: $(a, a_t) \leftarrow (\emptyset, -\infty)$, $(b, b_t) \leftarrow (\emptyset, -\infty)$} \\

\textsf{onStreamA(x: data, t: timestamp): }
\begin{enumerate}
    \item If b is not $\emptyset$, yield $(x, b)$
    \item If $t > a_t$, $(a, a_t) \leftarrow (x, t)$
\end{enumerate} 

\textsf{onStreamB(x: data, t: timestamp): }
\begin{enumerate}
    \item If a is not $\emptyset$, yield $(a, x)$
    \item If $t > b_t$, $(b, b_t) \leftarrow (x, t)$
\end{enumerate}
    }
}

Figure~\ref{fig:data-triggered-join} is an example of a data-triggered join.
Again for simplicity, we assume event time and processing time are the same. The join results are as follows:
join 1 (A1, B1, C1, \underline{D1});
join 2 (A1, \underline{B2}, C1, D1);
join 3 (A1, B2, C1, \underline{D2});
join 4 (A1, \underline{B3}, C1, D2);
join 5 (A1, B3, \underline{C2}, D2);
join 6 (\underline{A2}, B3, C2, D2).
In this way, we ensure that the system reacts immediately to new data at the expense of more frequent joins.
Data-triggered join is preferred when at least one data stream is bursty, as it is difficult to set a good time window with bursty data involved.

\subsection{When Is Data-triggered Join Better}
Data-triggered joins are more suitable for event-based streams whereas time windows are good aggregators for continuous data streams (e.g. sensor data).
In certain cases such as activity recognition, there is no activity of interest for most of the time.
For example, a Nest Cam only emits data when it finds people, vehicles, or animals in sight.
Data-triggered joins can capture these events as soon as they happen.
It is possible to combine time-triggered join and data-triggered join to get the best of two worlds.
\S\ref{sec:target-pred-freq} describes a hybrid method that primarily operates on a data-triggered basis, while strategically integrating time-triggered elements as a rate limiter.
Data-triggered joins further provide a completeness guarantee to the downstream data consumer. 
\emph{Every message is guaranteed to be present in at least one join tuple.}


In both time-triggered and data-triggered joins, we see repeated data appearing in multiple join results due to the lower frequency of some streams. Ideally, we want to send at most one copy of the same data over the network to avoid unnecessary bandwidth usage, especially if the data payload is large. Section~\ref{sec:lazy} proposes a novel technique called lazy data routing to address this problem.


\section{Lazy Data Routing}\label{sec:lazy}
The message broker system consists of a leader that orchestrates the entire message flow and multiple producers/consumers as message endpoints.
Data streams as producers publish data to the leader, and models as consumers consume data from the leader.
With this architecture, the leader can quickly become a point of contention since it has to process all the messages from/to all the different nodes.
Furthermore, large message payloads (e.g., images) can lead to a crucial networking bottleneck at the leader, as message broker systems are not designed to handle large messages.

\sys uses a novel messaging protocol to efficiently transfer data between nodes without placing an undue burden on the leader.
A message sent to the leader only contains message headers: a timestamp and a global source path.
The actual message payloads are not transferred; instead, they are kept and indexed on the node that collected the data.
A model subscribes to the topic and reads the headers as they come in.
If it wants a particular data payload, it retrieves that data lazily from the source node.

\begin{figure}[t]
    \centering
    \includegraphics[width=0.8\columnwidth]{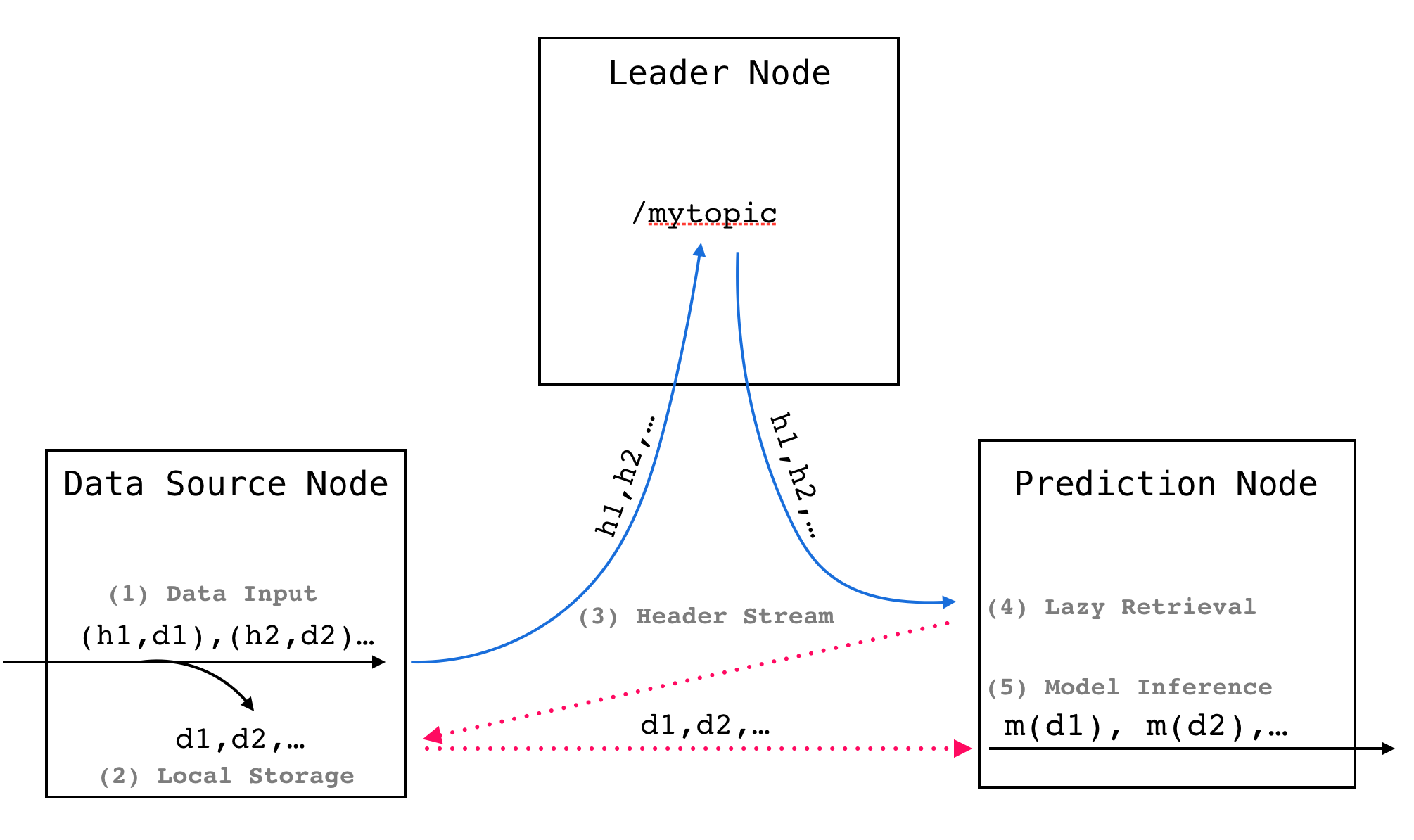}
    \caption{A figure illustrating the order of operations in the lazy data routing system used by \sys.}
    \label{fig:lazy}
\end{figure}

Figure~\ref{fig:lazy} illustrates this protocol. When collecting data, every data item added to a \texttt{DataStream} is annotated with a header (Figure \ref{fig:lazy}-1). We can think of this as a stream of $(h,d)$ tuples (header and data, respectively). After the tuple is created, the node locally writes the data to a time-indexed log (Figure \ref{fig:lazy}-2).
After this data is durably written, the header is published to the message broker on the leader (Figure \ref{fig:lazy}-3).
Nodes on the network can subscribe to streams of these headers.
Model inference requires the data payload, and that can be requested from the headers (Figure \ref{fig:lazy}-4). This data is transferred in a peer-to-peer fashion, and inferences happen over these streams (Figure \ref{fig:lazy}-5).

Lazy retrieval has a number of essential benefits for typical model-serving tasks. In general, these benefits are analogous to that of lazy computation.
First, many models predict at rates much slower than the rate of data collection. For example, a model that takes 30ms to evaluate can only process one example every 30ms. If the data collection rate is significantly faster than that, the model often has to downsample the input data. Lazy data routing allows us to avoid transferring the data payload to the leader in these cases. 
Next, this strategy reduces the size of the messages processed by the message broker reducing overheads in checkpointing and serialization/de-serialization.
As a result, we also find that it can enable increased parallelism as well.
Both of these benefits can be tied back to the traits of the machine learning setting mentioned above. (See \S\ref{sec:exp-lazy} for relevant experiments)

In certain cases, we allow users to force \sys to have eager message passing. Small messages, such as 1D arrays, can be transferred from data source nodes to worker nodes via the leader node. Essentially, this embeds the payload in the message headers.
In some networks, peer-to-peer communication is not available or not efficient.
We can default to eager message passing when needed to support these cases.

\section{Prediction Rate Control}\label{sec:rate-control}
Next, we show how to ensure this execution layer can meet particular model-serving service level objectives (SLOs). We leverage statistical approximations that exploit temporal similarity in typical data streams.
Every model in \sys is annotated with three timing parameters: (1) if it consumes multiple streams, a maximum tolerable \emph{skew}, (2)  a \emph{target prediction frequency}, which is an output rate limiter, and (3) a \emph{freshness threshold}, designed to discard stale messages originating from an earlier time.

\subsection{Message Skew}\label{sec:message-skew}
\sys gives the programmer an illusion of stream alignment, namely, streams associated with the same topic can be thought of as synchronized from the perspective of machine learning modeling. 
The consuming models receive tuples of headers corresponding to data from each of the sources.

Under the hood, \sys has to buffer streams locally to keep up this illusion.
The different data streams will arrive at different rates and have different system delays that cause misalignment.
We use a time interval-based interface for specifying alignment criteria.
Every topic has a maximum allowed time-skew (\S\ref{sec:streaming-inference}) between headers that can be produced.
Locally, the buffer retains a header until it receives matching header messages from other streams or the time-skew expires.
Thus, we can enforce a bounded-skew synchronization on the model side.
It is up to the user to set a reasonable time-skew limit for her specific task.
If the allowed time-skew is overly long, there is a risk of her encountering messages that lack proper synchronization.
Conversely, setting the time-skew limit too restrictively may result in the loss of some actually synchronized predictions owing to this stringent threshold.

\subsection{Hybrid Time- and Data-triggered Join}\label{sec:target-pred-freq}
The hybrid join in \sys is an innovative approach that combines the principles of both data-triggered and time-triggered joins. This method is designed to efficiently handle the challenges posed by high-velocity data streams and the processing capacity of models.

In essence, the hybrid join operates on two fundamental principles: rapid response to new data (inherited from the data-triggered join) and effective data management to avoid overloading the downstream model (inspired by the time-triggered join and backpressure mechanism~\cite{tassiulas1990stability}). When new data arrives from any stream, the hybrid join promptly triggers a joining operation, similar to a data-triggered join. This ensures that the system remains responsive to incoming data, allowing for timely processing and analysis.

However, to address the issue of data arriving faster than the downstream model can handle, the hybrid join incorporates a critical feature from the time-triggered approach: setting a minimum interval between consecutive processing instances, which we call \textit{target prediction frequency}. This interval acts as a throttle, ensuring that the downstream model is not overwhelmed by a continuous influx of data. If data arrives more rapidly than this set interval, the hybrid join mechanism will simply drop earlier data in the queue. This decision is based on the understanding that data delayed excessively in the queue may no longer be accurate or relevant for real-time decision-making.

By integrating these two approaches, the hybrid join offers a balanced solution that maximizes responsiveness to new data while maintaining a manageable processing load for the downstream model.

\subsection{Freshness Threshold}\label{sec:freshness-threshold}
As a streaming system, \sys prioritizes the timeliness of incoming data.
Stale data is essentially inaccurate data for latency-sensitive tasks.
Making use of such outdated information in real-time decision-making can lead to disastrous outcomes.
Our freshness threshold ensures the recency of data that the model can take.
It also acts as a rate limit when data arrives faster than the model inference rate.

\section{Experiments}\label{sec:exp}

\subsection{Experimental Setup}\label{sec:exp-setup}
All of our experiments are performed on a private ``edge cluster''.
Our hardware setup consists of 5 NVIDIA Jetson Nano Developer Kits, 4 Intel Skylake NUC computers, and a desktop PC. Each NUC is equipped with an Intel Core i3-6100U CPU, 16 GB RAM, and M.2 SSD.
The desktop PC features an Intel Xeon CPU E5-2603 v4 CPU, NVIDIA Quadro P6000 GPU, 64 GB RAM, and HDD.
Direct peer-to-peer connection is available between all nodes via 1Gbps Ethernet.
Throughout these experiments, we vary the network topology to test various scenarios.
In some experiments, only partial nodes are used.
These variations will be explained in respective sections, but one NUC is always used as the leader node.

As a primary baseline, we have configured PyTorch distributed~\cite{pytorch-distributed} on our edge cluster, with Gloo as the distributed communication backend.
We have also implemented an eager data routing architecture similar to ROS~\cite{quigley2009ros} within our framework to understand the key design decisions.
ROS is widely used in the sensor and robotics communities and provides a centralized message broker service.
However, ROS does not support lazy data routing, distributed stream synchronization, and adaptive rate control.
Additionally, we implemented a time-triggered join strategy similar to Apache Flink~\cite{flink}.
We also set up a local NTP server to make sure all nodes share a global wall clock time.

We borrow evaluation metrics from the streaming literature and a detailed description of these metrics is in Appendix~\ref{sec:exp-metrics}.

\subsection{Application: Human Activity Recognition}\label{sec:exp-opportunity}
\noindent \textbf{Description. } We use the Opportunity dataset for human activity recognition~\cite{opportunity-dataset,opportunity-challenge} as an example.
Data from multiple motion sensors were collected about every 33ms while users were executing typical daily activities. For each subject, there are five activity of daily living (ADL) runs, and each run lasts 15-30 minutes.
We take the first subject's first four ADL runs as the training set and the last ADL run as the test set. When played at 2x speed, the last ADL run takes 8 minutes and 22 seconds.
We partition the first 134 columns vertically into four disjoint subsets, each placed on one of four nodes (3 NUCs and 1 Jetson Nano) as data sources.
The subsets are distributed as follows: columns 1-37 (accelerometers), 38-76 (IMU back and right arm), 77-102 (IMU left arm), and 103-134 (IMU shoes).
We train an aggregated random forest model with scikit-learn~\cite{scikit-learn} for all 134 features as an early fusion baseline, and also four separate random forest models for each subset of features to evaluate an ensemble-based late fusion method.
We primarily evaluate \sys with the late fusion deployment: one RF model at each data source node and we ensemble local predictions at another node.
This simulates a scenario where there is a small amount of compute on each wearable sensor and that compute is used to reduce the data communicated to make a global prediction across those sensors.
Due to space limits, we defer the early vs. late fusion discussion to Appendix~\ref{sec:appendix-late-fusion}.

In our best-effort PyTorch implementation, we use the \texttt{gather()} API to aggregate data from multiple data source nodes.
PyTorch distributed requires all tensors to be the same size to be gathered, so we have to pad each local tensor to the maximum size with zeros.
The individual streams are fast enough that misalignments can occur due to queueing delays.
However, PyTorch enforces that multiple data sources are always perfectly synchronized, as it does not begin the actual computation until data from all data sources has been gathered, and it only gathers new data after finishing previous predictions.
Such strict requirement does not exist in \sys, as we are able to set a reasonable skew (\S\ref{sec:message-skew}) in \sys.

\vspace{0.25em} \noindent \textbf{Queueing in \sys is better suited for real-time applications. }
\begin{figure}[t]
    \centering
    \begin{minipage}{0.48\columnwidth}
        \centering
        \includegraphics[width=\textwidth]{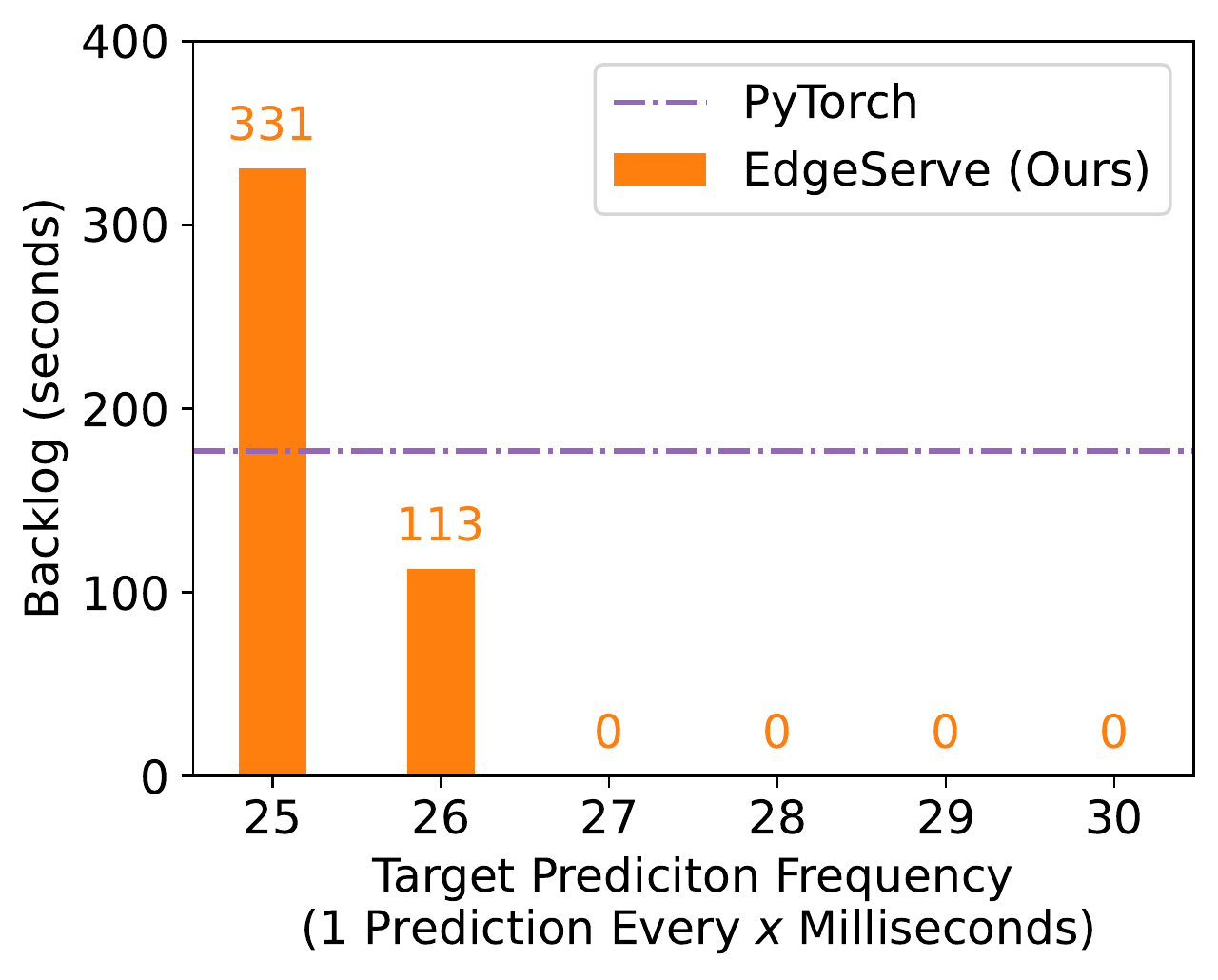}
        \caption{Measure of backlog in the activity recognition task. More frequent predictions are on the left side.}
        \label{fig:exp-opportunity-latency}
    \end{minipage}\hfill
    \begin{minipage}{0.48\columnwidth}
        \centering
        \includegraphics[width=\textwidth]{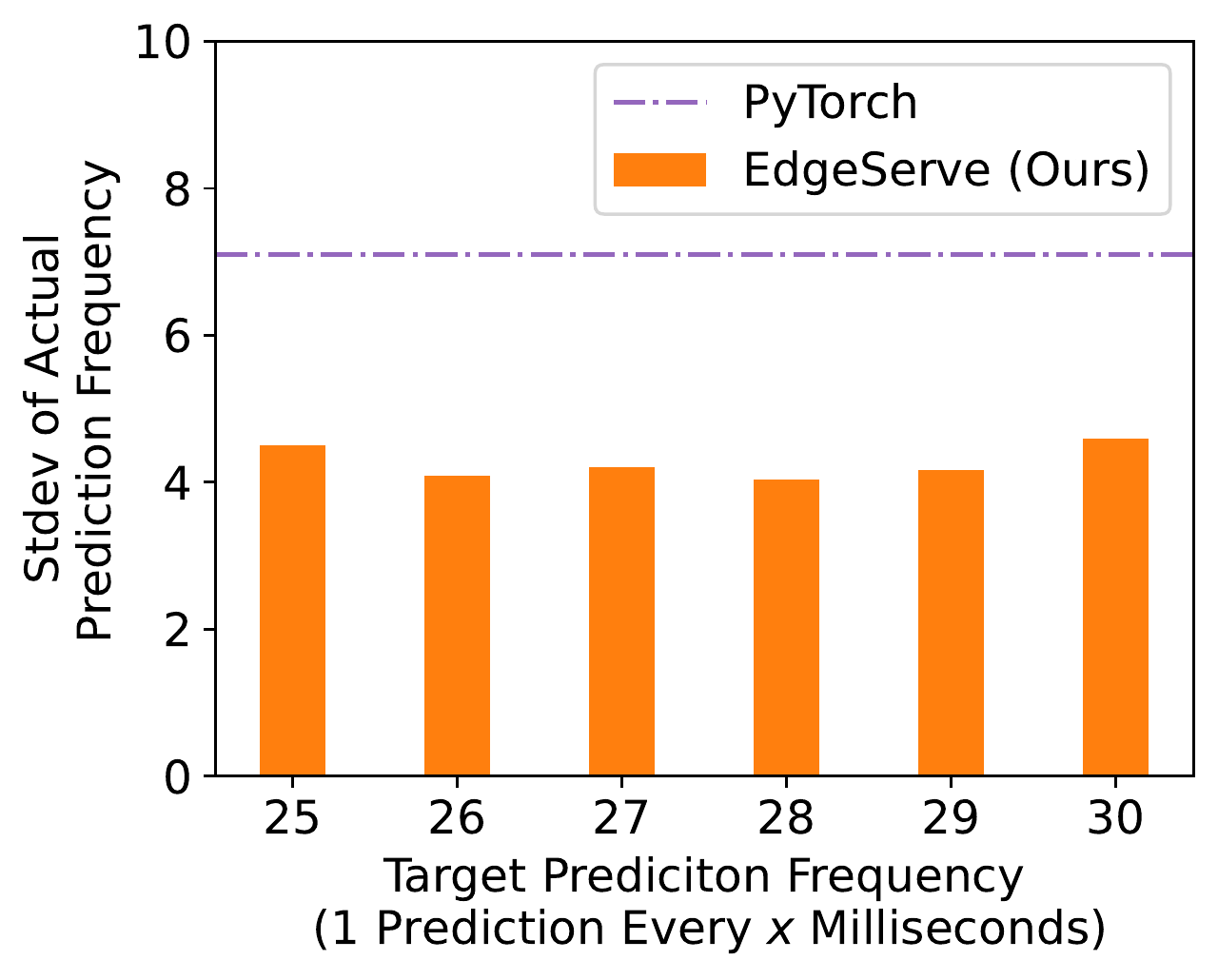}
        \caption{Standard deviation of actual prediction frequency, where \sys maintains a lower variability.}
        \label{fig:exp-opportunity-stdev}
    \end{minipage}
\end{figure}
First, we evaluate the ability of the system to even issue real-time predictions by measuring the \textit{backlog} in the system, or the accumulated queuing time as defined in Appendix~\ref{sec:exp-metrics-backlog}.
Unlike PyTorch, \sys deployments have a prediction frequency target and can use this target to automatically downsample data to meet real-time requirements.
We illustrate the improvements in Figure~\ref{fig:exp-opportunity-latency}. The x-axis is the target prediction frequency (\S\ref{sec:target-pred-freq}) designated by the end-user, where a larger number means a lower frequency; the y-axis is the \textit{backlog} for each of the serving systems over this dataset.
The compute part of the task itself takes about 23ms to complete, and a near-zero number in backlog means the inference is processed in real time.
\sys offers a no-backlog queue for a wider range of prediction frequency targets ($\geq 27$ms/pred).
However, a long queue of unprocessed examples is quickly developed without proper rate control (e.g. when target prediction frequency $\leq 26$ms per prediction).
Since PyTorch lacks a message queue and rate control, it has to process each example individually and trigger joins in a strictly synchronous manner, leading to an unsatisfactory backlog.

\begin{figure}[t]
    \centering
    \begin{minipage}{0.48\columnwidth}
        \centering
        \includegraphics[width=\textwidth]{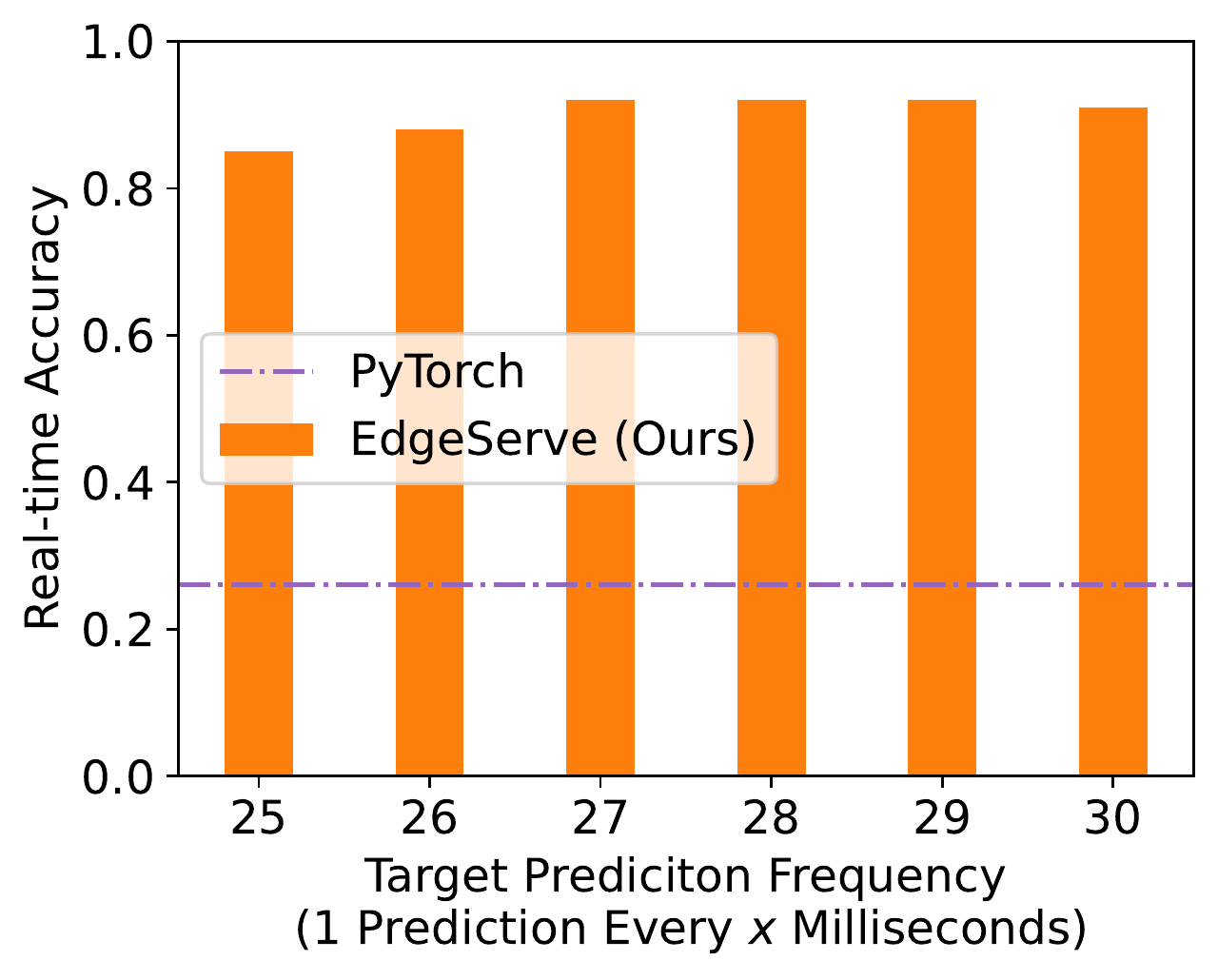}
        \caption{Overall real-time accuracy for human activity recognition task measured in F-1 score. }
        \label{fig:exp-opportunity-accuracy}
    \end{minipage}\hfill
    \begin{minipage}{0.48\columnwidth}
        \centering
        \includegraphics[width=\textwidth]{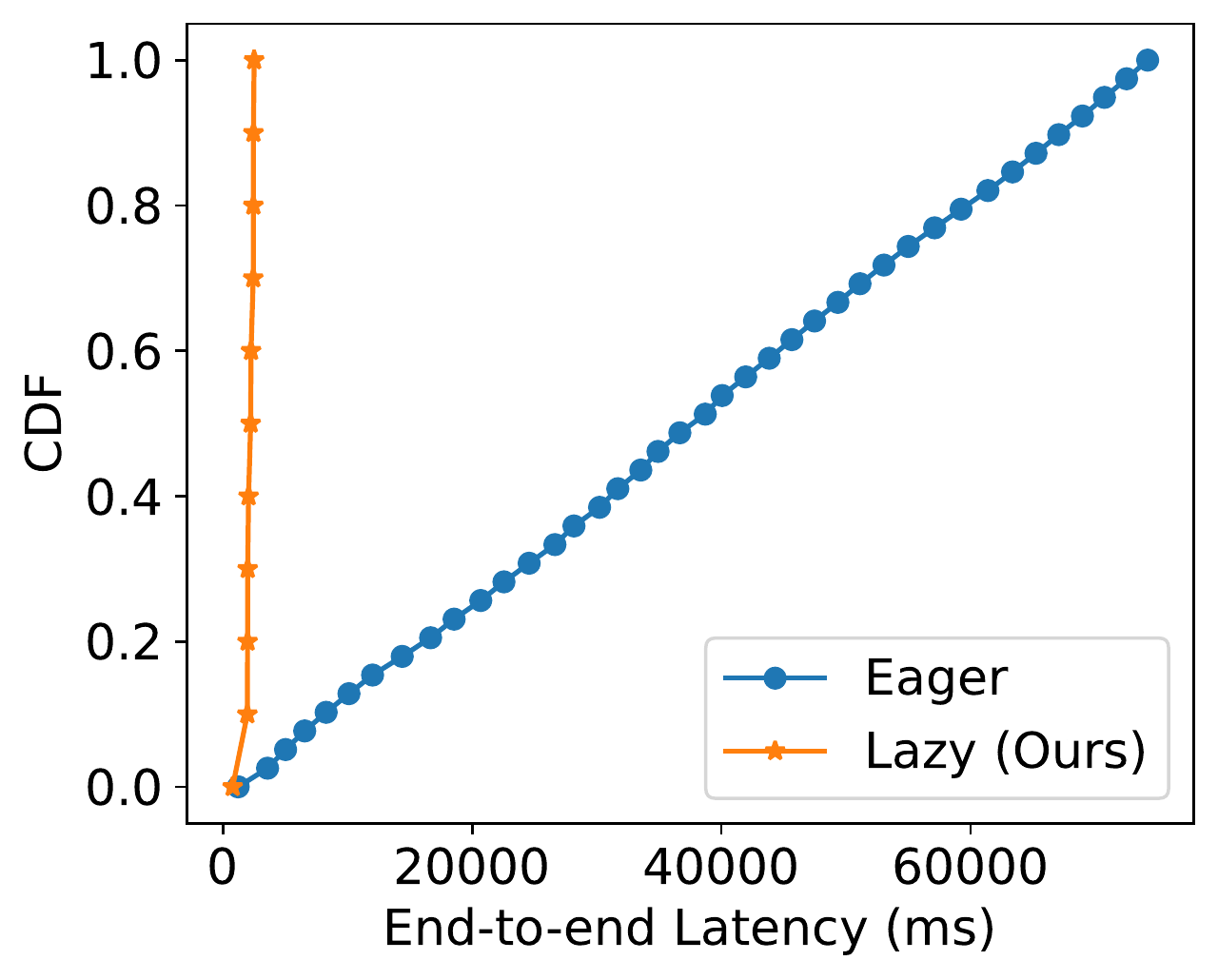}
        \caption{CDF of end-to-end latency for eager data routing vs. lazy data routing.}
        \label{fig:exp-nuscenes-e2e-latency}
    \end{minipage}
\end{figure}

Even if PyTorch could meet real-time prediction targets, we find that the variability in prediction latencies is quite high.
In Figure~\ref{fig:exp-opportunity-stdev}, we see a much higher variability in actual prediction frequencies for PyTorch than \sys across all user-defined rates.
This is because PyTorch communicates in a synchronous fashion, and has to account for the variability of all 4 nodes making local predictions with local data streams.

\vspace{0.25em} \noindent \textbf{Queueing Delays Reduce ``Real-Time'' Accuracy. }
In real-time serving scenarios, the timeliness of predictions becomes a key concern. For latency-sensitive tasks, a delayed prediction equates to an incorrect one. To evaluate the timeliness of predictions, we introduce \textit{real-time accuracy} as a measure, which evaluates the accuracy of predictions against the most recent label at the time of prediction. For instance, if a prediction is made between two consecutive labels at times $t_1$ and $t_2$ ($t_1 \leq t_2$), its accuracy is compared with the label at $t_1$.
Since we assume adjacent examples are likely similar, we expect roughly correct prediction results when the examples arrive slightly late. However, if the examples arrive significantly late, they are likely outdated and yield incorrect predictions.


Figure~\ref{fig:exp-opportunity-accuracy} shows the real-time accuracy of \sys and PyTorch under various target prediction frequencies.
PyTorch distributed is not able to issue accurate predictions because data is communicated in a synchronous manner. It is unable to downsample the input stream even if the node is overloaded, making most of its predictions outdated.
In contrast, \sys, at a target prediction frequency of 25ms, experiences a greater backlog compared to PyTorch but achieves superior real-time accuracy. This advantage is primarily due to the experiment setup of 3 NUCs and 1 Jetson Nano for local model inference. The NUCs process the CPU model more efficiently than the Jetson Nano, resulting in a significant portion of the backlog being attributed to the Jetson Nano, as it completes local inference later than the NUCs. This situation leads to a notable message skew. To mitigate this, \sys selectively skips data that exceeds the maximum tolerable skew (\S\ref{sec:message-skew}). This strategy of skipping mostly inaccurate data significantly boosts \sys's real-time accuracy.
Furthermore, when the target prediction frequency is set above 26ms/pred, \sys sees less backlog and achieves even higher real-time accuracy.
This improvement results from \sys's capability to instantly process fresher data that, while not perfectly synchronized, falls within an acceptable time skew.

\subsection{Application: Autonomous Driving}\label{sec:exp-nuscenes}
We use a subset of the nuScenes self-driving dataset for autonomous driving~\cite{nuscenes} consisting of 6 cameras and a lidar sensor.
All cameras generate 10 frames per second and the lidar sensor emits at 2 Hz.
Each camera is connected to a separate NVIDIA Jetson Nano running pre-trained YOLOv5n model~\cite{yolov5} on GPU.
The lidar sensor is connected to a NUC node, which preprocesses the data and then transfers it to our desktop PC equipped with NVIDIA Quadro P6000 GPU running pre-trained CenterPoint model~\cite{yin2021center}.
Communication incurs considerable cost here as preprocessed lidar data is very large.
All predictions are sent to another NUC node, which triggers the join and yields synchronized predictions.

\begin{table}[]
\small
\begin{tabular}{ll|lll}
\multicolumn{2}{l|}{\makecell{Queueing time\\ (ms)}} & \makecell{Time-triggered\\ (Flink-like)} & \makecell{Data-triggered\\ (Ours)} & Speedup \\ \hline
\multirow{2}{*}{Lidar}       & Med.    & 463.60         & 16.56          & 28.00x  \\
                             & P95       & 963.20         & 48.35          & 19.92x  \\ \hline
\multirow{2}{*}{Cam 1}    & Med.    & 51.04          & 5.64           & 9.06x   \\
                             & P95       & 126.06         & 13.56          & 9.30x   \\ \hline
\multirow{2}{*}{Cam 2}    & Med.    & 64.05          & 9.59           & 6.68x   \\
                             & P95       & 124.08         & 18.27          & 6.79x   \\ \hline
\multirow{2}{*}{Cam 3}    & Med.    & 56.62          & 9.76           & 5.80x   \\
                             & P95       & 143.45         & 17.84          & 8.04x   \\ \hline
\multirow{2}{*}{Cam 4}    & Med.    & 74.30          & 5.23           & 14.19x  \\
                             & P95       & 117.08         & 13.08          & 8.95x   \\ \hline
\multirow{2}{*}{Cam 5}    & Med.    & 45.32          & 7.63           & 5.94x   \\
                             & P95       & 88.87          & 15.90          & 5.59x   \\ \hline
\multirow{2}{*}{Cam 6}    & Med.    & 57.33          & 5.38           & 10.66x  \\
                             & P95       & 109.61         & 13.20          & 8.31x   \\
\end{tabular}
\caption{Queueing time for time- vs. data-triggered joins.}
\label{tab:exp-nuscenes-joins}
\end{table}

First, we compare the end-to-end latency between eager and lazy data routing, applying data-triggered join in both scenarios.
In the lazy data routing approach, we implemented a freshness threshold SLO (\S\ref{sec:freshness-threshold}) that discards data older than 500ms.
As depicted in Figure~\ref{fig:exp-nuscenes-e2e-latency}, the CDF of end-to-end latency demonstrates that lazy data routing significantly reduces latency by only pulling data with recent timestamps.
This reduction is particularly notable since communication is the primary bottleneck in this task.
Notably, lazy data routing led to the skipping of 72.5\% of predictions that failed to meet our freshness threshold SLO compared to eager data routing.

Second, we compare the queueing time (as defined in~\ref{sec:exp-metrics}) between time-triggered and data-triggered joins, applying lazy data routing in both scenarios.
For time-triggered join, we set the time interval of joins to be every 1 second.
For data-triggered join, we issue a join as soon as a new example that meets our freshness threshold SLO comes in.
Table~\ref{tab:exp-nuscenes-joins} shows the median and 95th percentile of queueing time for each data source.
Data-triggered join reduces the queueing time by up to 28x as it does not have to wait for fixed intervals.

\subsection{Application: Network Intrusion Detection}\label{sec:exp-network}
\sys natively allows multiple producers and multiple consumers to operate on the shared message queue at the same time, which is an essential communication paradigm in decentralized prediction but not currently supported by PyTorch or TensorFlow.
We use a public Network Intrusion Detection dataset from Canadian Institute for Cybersecurity (CIC-IDS2017)~\cite{cic-ids2017} and an existing model~\cite{kostas2018} to differentiate malicious traffic from benign network traffic.
Specifically, we partition the data horizontally into four disjoint subsets by ``Source IP'' for our four data source nodes. The underlying assumption is that network traffic from different source IP addresses may be collected separately.

If a web attack is detected, the related network packet needs to be sent to a specific destination node, but the actual computation can be done anywhere. 
We show that \sys can support three deployment strategies: (Early fusion, topology 1) transfer all data to the prediction node that does all computations in a centralized way; (Early fusion with parallelism, topology 2) transfer all data from data source nodes to an intermediate shared queue, where four prediction nodes can pull data from when they become available, and they need to inform the destination node if an attack is detected; or (Late fusion, topology 3) data source nodes do computations locally and only transfer data to the destination node if an attack is detected.  

In an early fusion setting, PyTorch distributed is able to process 41.94 examples per second, while \sys can process 47.58 examples per second.
This is the baseline setting of both systems, and the performances of both systems are comparable.
In an early fusion with parallelism setting that is only supported by \sys, thanks to its queuing design, 182.57 examples are processed per second, which is almost a linear (3.84x) speedup compared to a centralized setting given that we now have 4 prediction nodes.
In a late fusion setting, we make all 4 data source nodes also local prediction nodes, and PyTorch achieves 181.33 examples per second while \sys takes 197.30 examples per second. For both systems, superlinear speedup (4.32x and 4.15x compared to centralized, respectively) is achieved by making the most of local computational resources and communicating only local prediction results instead of the entire dataset.
Since we use the same model/sub-models for both \sys and PyTorch without synchronization issues, the accuracies of predictions between both systems are the same.

\subsection{Micro-benchmarks}

\subsubsection{Data-triggered Joins Are More Responsive}\label{sec:exp-joins}
\begin{table}[t]
\begin{tabular}{l| l|l}
\multicolumn{1}{c|}{\multirow{2}{*}{Join strategy}}
    & \multicolumn{2}{c}{Reaction time} \\
    & Median &P95\\ \hline
Data-triggered & 9.02ms & 10.28ms\\
Time-triggered (time window: 1s) & 0.5s & 0.9s\\
Time-triggered (time window: 5s) & 2.5s & 4.7s\\
\end{tabular}
\caption{Reaction time for data- and time-triggered joins.}
\label{tab:exp-reaction-time}
\end{table}
This micro-benchmark evaluates the responsiveness between time-triggered join and data-triggered join.
We use two NUC nodes as data sources, one NUC node as the message broker, and one more NUC node performing the join.
We show how data-triggered joins significantly reduce reaction time for latency-sensitive applications.
We have a steady stream that arrives every 5 seconds (5 MB each) and a bursty stream that arrives at 10 Hz (1 Byte each) for a minute.
Real-time decision-making requires the latest information from both streams.
The \textit{reaction time} (as defined in Appendix \ref{sec:exp-metrics-latency}) for both join strategies during that one minute is shown in Table~\ref{tab:exp-reaction-time}.
Data-triggered join achieves a much better reaction time without the difficulty of setting a reasonable time window.

\subsubsection{Benefits of Lazy Data Routing}\label{sec:exp-lazy}
In \S\ref{sec:lazy}, we described our lazy data routing model as an alternative to a ROS-like system that eagerly transfers raw data through a centralized broker.
Now, we evaluate the pros and cons of the lazy data routing model.
We employ one NUC node as the data source and one NUC node as the receiver in this subsection, except for the parallelism experiment where the number of receiver nodes is varied.


\begin{figure}[t]
    \centering
    \includegraphics[width=0.9\columnwidth]{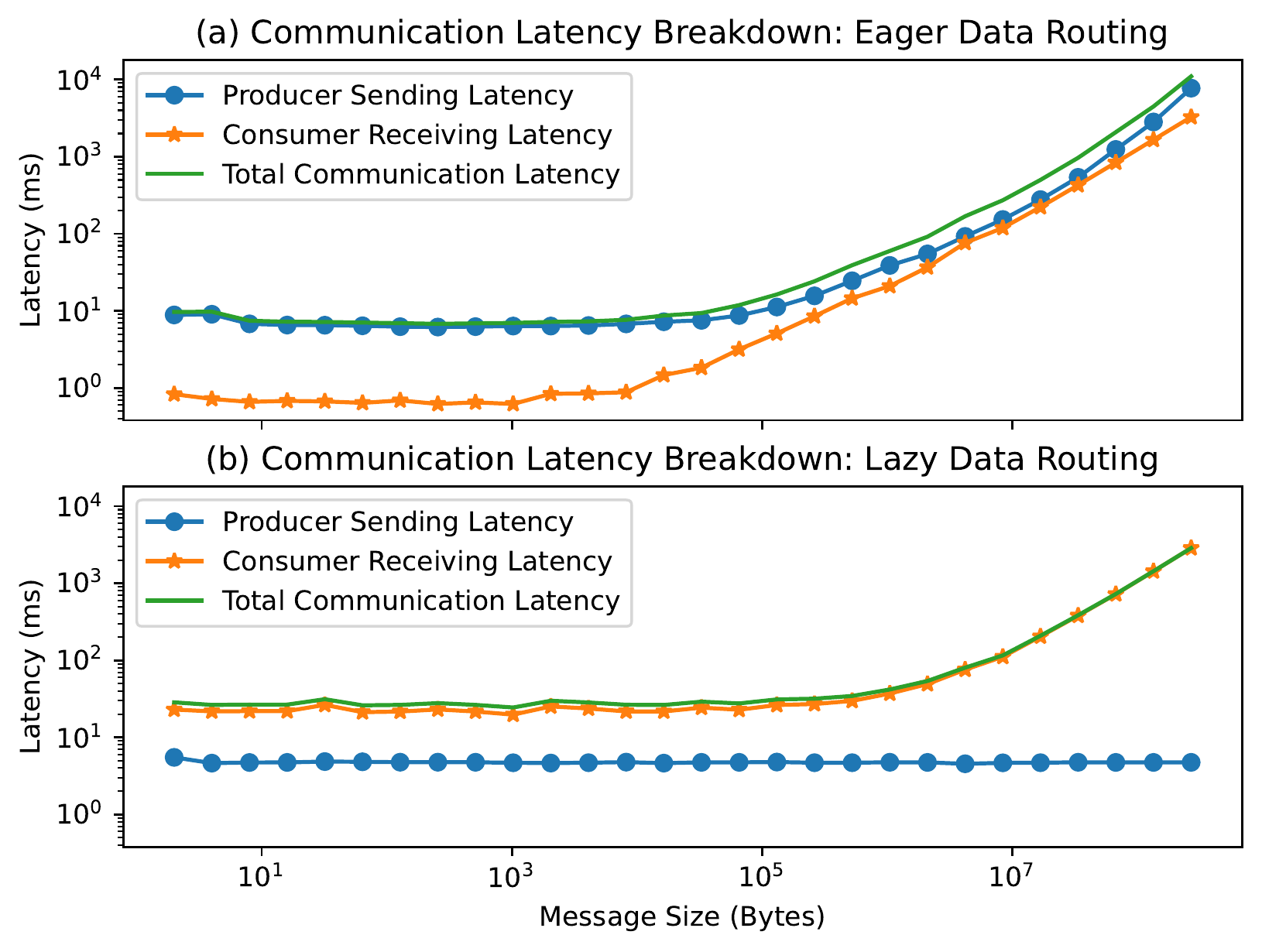}
    \caption{Lazy data routing reduces latency on producer side but has fixed overhead on consumer side. Both axes are log-scaled.}
    \label{fig:exp-comm-latency-breakdown}
\end{figure}

\begin{figure}[t]
    \centering
    \includegraphics[width=0.9\columnwidth]{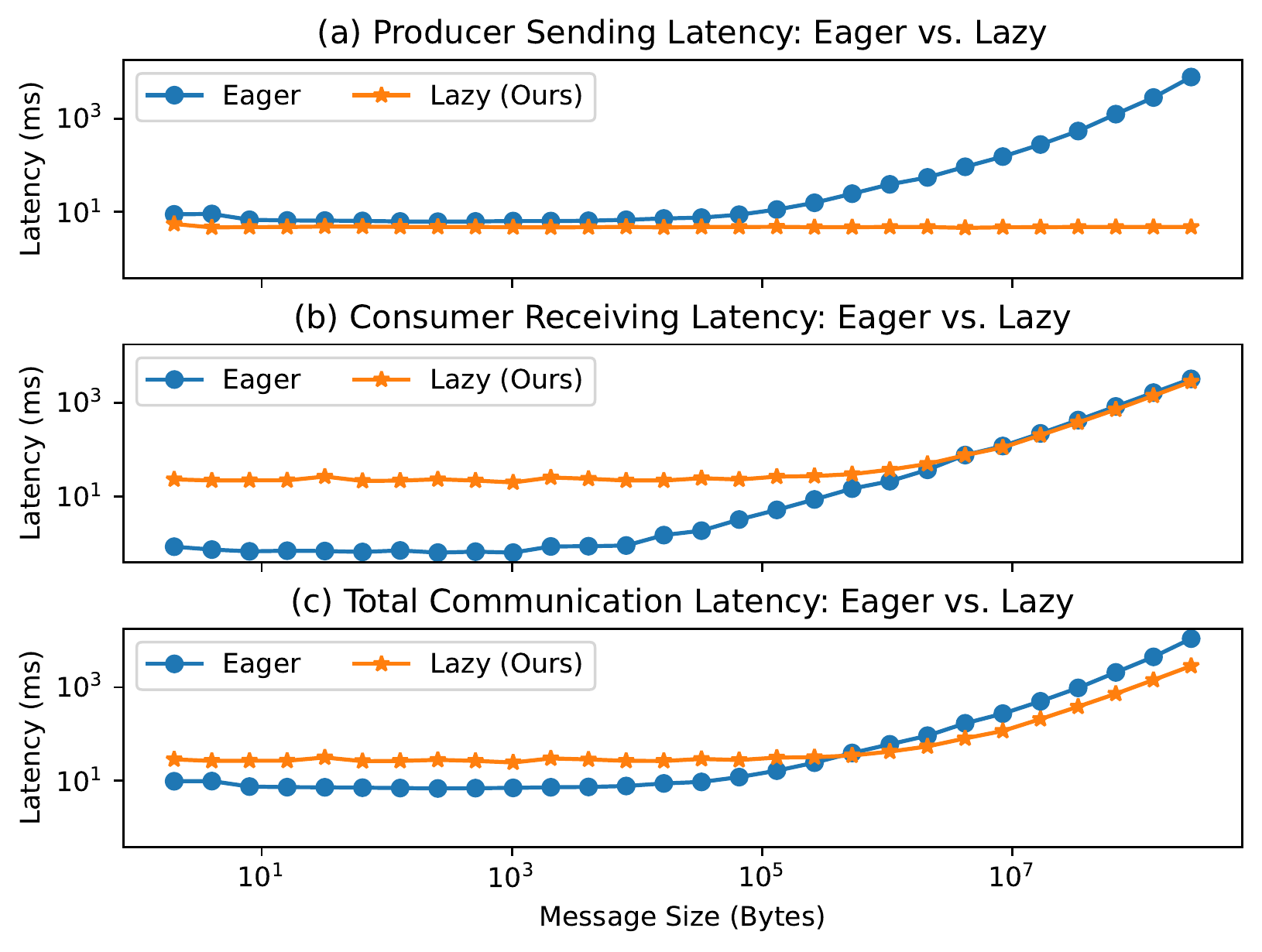}
    \caption{Lazy data routing reduces latency on producer side but has fixed overhead on consumer side. Both axes are log-scaled.}
    \label{fig:exp-comm-latency-eager-lazy}
\end{figure}

First, we send a series of messages of different sizes from a data source node to a receiver node, through the leader node. No actual computation is performed. We compare the communication latency between eager and lazy data routing.

\vspace{0.25em} \noindent \textbf{Lazy Data Routing Reduces Latency on Producer but Has Fixed Overhead on Consumer.}\label{sec:exp-lazy-tradeoff}
Since we only need to transfer the headers instead of raw data in our lazy data routing model, the latency on producer side remains a negligible number even if the message is huge.
As shown in Figure~\ref{fig:exp-comm-latency-breakdown}a and~\ref{fig:exp-comm-latency-eager-lazy}a, a ROS-like eager data routing model could result in very high latency when sending large messages, which itself could force subsequent messages to queue up and become outdated when they arrive.
In contrast, our lazy data routing model makes sure that message headers are sent in milliseconds, which never blocks the rest of messages. The consumers may choose to downsample some data and only fetch necessary data.

Whenever the consumer needs to fetch raw data, there is a fixed overhead to establish P2P connections even if the actual data is just a few bytes. This fixed overhead can be amortized when the actual data is larger, as depicted in Figure~\ref{fig:exp-comm-latency-breakdown}b and~\ref{fig:exp-comm-latency-eager-lazy}b.

In summary, lazy data routing is more performant when the messages transferred are larger in size.
As shown in Figure~\ref{fig:exp-comm-latency-eager-lazy}c, eager data routing actually has a lower total communication latency when the messages are smaller than 512KB in size, and lazy data routing performs better when the messages are larger than 512KB in size.

\begin{figure}[t]
    \centering
    \begin{minipage}{0.48\columnwidth}
        \centering
        \includegraphics[width=\textwidth]{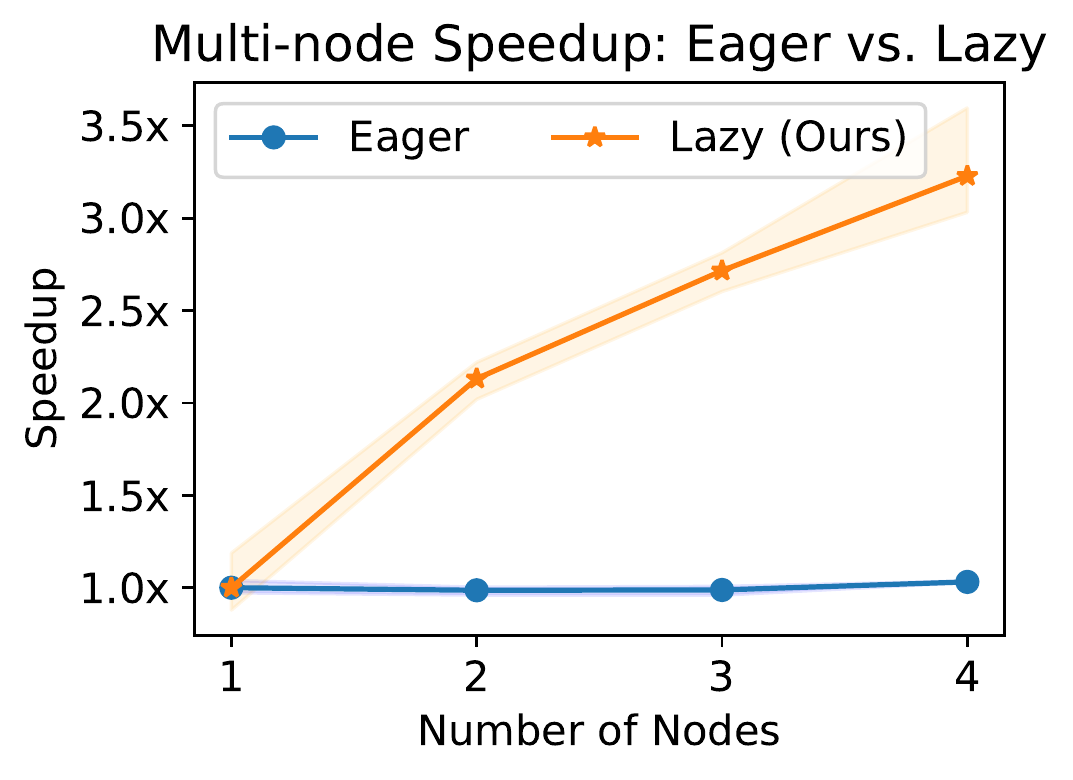}
        \caption{Lazy data routing scales out well while eager data routing does not.}
        \label{fig:exp-speedup}
    \end{minipage}\hfill
    \begin{minipage}{0.48\columnwidth}
        \centering
        \includegraphics[width=\textwidth]{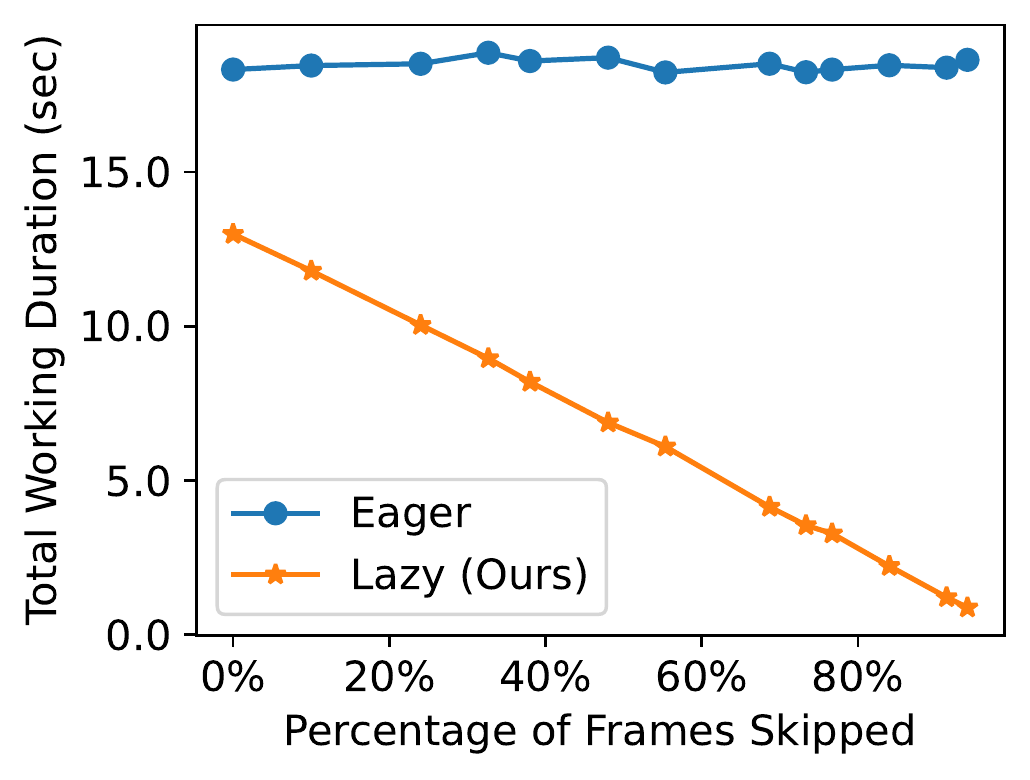}
        \caption{Lazy data routing saves communication when some data is skipped.}
        \label{fig:exp-lazy-skipped}
    \end{minipage}\hfill
\end{figure}

\vspace{0.25em} \noindent \textbf{Lazy Data Routing Naturally Supports Parallelism.}\label{sec:exp-lazy-parallelism}
It is very common for multiple consumers to fetch data from one or more producers at the same time.
In our lazy data routing model, since messages are transferred in a peer-to-peer fashion, the leader node only has a very light workload to process tiny headers simultaneously, saving precious bandwidth at the leader node.
However, in the eager data routing model, the leader node can be blocked when a piece of large message is going through the leader node from a producer to a consumer. As a result, other producers and/or consumers running in parallel cannot send or receive messages at the same time.

To compare the scalability of communication between eager and lazy data routing models, we have one producer continuously sending the same 512KB message for a total of 100 times to a shared queue. We gradually increase the number of consumer nodes from 1 to 4 and see how it scales out. While no actual computation is done, we measure the total working duration and use the single-node setup for both the eager and lazy data routing as the 1.0x baseline.
Figure~\ref{fig:exp-speedup} shows how the eager data routing model fails to scale out with more consumer nodes, while our lazy data routing model achieves reasonable speedup. The line shadows represent the lower and upper bounds of repeated experiments.

\vspace{0.25em} \noindent \textbf{Lazy Data Routing Performs Better with Network Contention.}\label{sec:exp-congestion}
Our lazy data routing model is especially beneficial when the leader node is busy with network requests.
We specifically construct a task where the message payload is large: real-time inference over video streams.
In this experiment, two webcams capture the same moving QR code from different positions. 
Both videos are 150 frames long at 1920x1080 resolution.
Each camera is connected to a unique data source node on the network.
For multi-camera tracking, the QR code has to be detected in both streams and corresponded in time-aligned frames from both cameras.
So these two data streams need to be joined at the prediction node.
We simulate a congestion scenario where the network bandwidth at the leader node is limited.
Note that the rest of the network retains its full speed; the only congestion is at the leader node.

Table~\ref{tab:congestion} shows the results.
With no congestion, the system can process roughly 0.8 frames per second in both lazy and eager data routing. With congestion, the story is very different. Our lazy data routing is tolerant, while transferring raw frames in a ROS-style eager communication pattern can be extremely slow when the network is congested. The total working duration increases by a factor of 7 simply due to congestion. Without care, distributed, multi-sensor deployments can easily lose real-time processing capabilities if the broker becomes a point of contention. These experiments illustrate the value of \sys in a controlled scenario, where we can isolate performance differences.

\begin{table}[t]
\centering
\begin{tabular}{l|l|l}
Data routing strategy & Rate limit (up/down) & Time    \\ \hline
Lazy (ours) & No limit                       & 3m 10s  \\
Lazy (ours) & 1 Mbps / 1 Mbps                & 3m 12s  \\
Eager (similar to ROS)       & No limit                       & 3m 16s  \\
Eager (similar to ROS)         & 20 Mbps / 20 Mbps              & 21m 32s
\end{tabular}
\caption{Total working duration with network bandwidth limits.}
\label{tab:congestion}
\end{table}

\vspace{0.25em} \noindent \textbf{Lazy Data Routing Performs Better with Data Skipping.}\label{sec:exp-lazy-data-skipping}
Apart from network congestion, lazy data routing is also valuable when data skipping is employed to ensure the timeliness of prediction results.
We take one of the two 150-frame videos mentioned earlier and transfer these frames from one node to another, via the leader node. Each frame is about 6 MB in its uncompressed form. No actual computation is performed as we are only interested in the communication cost.
In Figure~\ref{fig:exp-lazy-skipped}, we illustrate how much communication cost can be saved by lazy data routing.
On the x-axis, we have a variable percentage of frames skipped due to adaptive rate control described in \S\ref{sec:rate-control}; on the y-axis, we measure the total working duration defined in \S\ref{sec:exp-metrics-latency}.
Even when no frames are skipped, our lazy data routing model performs better than the eager data routing model due to the eliminated overhead of transferring a large amount of data through the leader node.
When more frames are skipped, our lazy data routing saves communication time almost linearly to the number of frames skipped by the downstream node.
On the other hand, the ROS-style eager routing pattern spends roughly the same time on communication even if most of the frames are skipped by the downstream model, because it would transfer the entire data payload upfront anyway.

\subsection{Comparison with Ray Serve}\label{sec:exp-system-overhead}
We conduct an object detection task with another serving system, Ray Serve~\cite{ray}, with a sample of nuScenes~\cite{nuscenes} camera data and pre-trained YOLOv5n model~\cite{yolov5} on a single NVIDIA Jetson Nano.

Single-node performance between Ray Serve, \sys, and an ideal case where the job runs locally without any communication is presented in Table~\ref{tab:exp-nuscenes-overhead}.
First, we enforce the freshness threshold SLO (\S\ref{sec:freshness-threshold}) of 1 second and see how many examples must be skipped in order to hit the SLO.
Since the data comes faster than the model's inference speed, 19.0\% of incoming data has to be skipped even in an ideal case.
\sys skips a bit more examples than ideal, but the overhead is reasonably small.
Ray Serve, however, skips 89.4\% of incoming data, which means the system consumes more computational resources than the task itself.
Second, we drop the SLO requirement and see how long it takes for each system to complete the task without downsampling.
In an ideal scenario, the model runs for 25 seconds to finish the dataset.
\sys spends 26 seconds, which presents negligible system overhead.
Ray Serve, on the other hand, spends 2m40s finishing the task, which is 6.4x slower compared to ideal due to its complex design.
In addition, we compare key design decisions between Ray Serve and \sys in Table~\ref{tab:rayserve-comp}.

\begin{table}[]
\small
\centering
\begin{tabular}{l|l|l|l}
    & Ray Serve  & \sys & Ideal \\ \hline
\makecell{\% examples skipped\\ (w/ SLO enforced)}             & 89.4\%   & 22.2\% & 19.0\%     \\ \hline
\makecell{Total working duration\\ (w/o SLO enforced)}         & 2m40s    & 26s    & 25s
\end{tabular}
\caption{Single-node performance of Ray Serve, \sys, and an ideal case.}
\label{tab:exp-nuscenes-overhead}
\end{table}

\section{Related Work}
\noindent \textbf{Model serving.}
Current machine learning model serving systems, including Clipper~\cite{clipper}, TensorFlow Serving~\cite{tfserving}, and InferLine~\cite{inferline}, all assume that the user has manually programmed all necessary data movement.
Recent systems have begun to realize the underappreciated problem of data movement and communication-intensive aspects of modern AI applications.
For example, Hoplite~\cite{hoplite} generates data transfer schedules specifically for asynchronous collective communication operations (e.g., broadcast, reduce) in a task-based framework, such as Ray~\cite{ray} and Dask~\cite{dask}.
However, they have yet to address the trade-offs in time-synchronization between different data sources when they do not arrive at the same time.

\begin{table}[t]
    \centering
    \small
    \begin{tabular}{c|c|c}
         & Ray Serve & \sys (ours) \\ \hline
        Communication & HTTP POST & Message queue \\ \hline
        Data routing & Eager & Eager / Lazy \\ \hline
        Downsampling & Not supported & Supported \\ \hline
        Result delivery & Back to caller & Can route to any topic \\ \hline
        In-out relationship & 1:1 request-response & 1:n/n:1 with joins \\
    \end{tabular}
    \caption{Design decisions: Ray Serve vs. \sys (ours)}
    \label{tab:rayserve-comp}
\end{table}

\noindent \textbf{Edge computing.}
On the other hand, there has been a steady trend towards moving model serving to resources closer to the point of data collection, or the ``edge''. 
The primary focus of model serving on the edge has been to design reduced-size models that can efficiently be deployed on lower-powered devices~\cite{han2015deep, edgedrive, videoEdge, distream}.
Simply reducing the computational footprint of each prediction served is only part of the problem, and these tools do not support data routing when the relevant features might be generated on different edge nodes.

\noindent \textbf{Distributed communication.}
The closest existing tools are those designed for distributed training of ML models. TensorFlow Distributed~\cite{TensorFlow}, for example, allows both all-reduce (synchronous) and parameter server (asynchronous) strategies to train a model with multiple compute nodes.
Another popular framework, PyTorch distributed~\cite{pytorch-distributed}, supports additional collective communication operations such as gather and all-gather with Gloo, MPI, and NCCL backends. One might ask, can we perform distributed inference using these existing distributed training frameworks? Technically it is possible, but as we have shown in \S\ref{sec:exp-opportunity}, the performance is unsatisfactory because such frameworks are optimized for maximum throughput but not end-to-end latency.

\noindent \textbf{Message queues.}
An alternative to distributed communication is message queues.
Existing message queues typically expose a publish-subscribe model that moves messages between services.
Compared with transient queues like RabbitMQ~\cite{rabbitmq} and direct TCP connections as used in Naiad~\cite{naiad}, log-based queues, such as Kafka~\cite{kafka} and Pulsar~\cite{apachepulsar}, ensure the order of messages and are persistent in nature. Users can always replay the logs for debugging purposes. 
Modern message services offered by cloud providers, such as Google PubSub and Amazon SQS, generally have higher latencies in the hundreds of milliseconds due to synchronous data replication across multiple zones.


\noindent \textbf{Dataflow systems.}
Existing batch processing and stream processing systems support dataflow computations over a dataflow graph~\cite{goog_data,spark,spark-structured-streaming,naiad,apache-samza,apachestorm,twitter-heron,TensorFlow,pytorch,flink}.
These systems are optimized for windowed operations over unbounded data, because a fixed frequency is preferred in typical data analysis workloads.
In contrast, model-serving applications have a much finer granularity of data input and more sensitivity in decision making. Fixed time-windowing is usually not suitable for bursty data. Decision making is based on up-to-date predictions as every single new event arrives.

\noindent \textbf{Temporal synchronization.}
Similarly, this problem is more than just a stream processing problem.
Traditional relational stream processing systems, e.g.,~\cite{chandrasekaran2003psoup}, have stringent requirements for temporal synchronization where they model such an operation as a temporal join. 
These systems will buffer data, indefinitely if needed, to ensure that corresponding observations are properly linked.
While desirable for relation query processing, this approach is excessive in machine learning applications which have to tolerate some level of inaccuracy anyway.
In addition, existing streaming join algorithms look for values from different streams with the same key~\cite{celljoin, splitjoin, DBLP:conf/sigmod/TeubnerM11, DBLP:journals/pvldb/RoyTG14}.
Multi-modal machine learning inference, however, pays more attention to the time skew between streams than the join predicate, as data sources might come at different rates.
In this setting, a looser level of synchronization would benefit the system and improve performance.

In the context of sensing, ROS (Robot Operating System)~\cite{quigley2009ros} is an open-source framework designed for robotics research. It incorporates an algorithm called \texttt{ApproximateTime} that tries to join messages coming on different topics at different timestamps. This algorithm can drop messages on certain topics if they arrive too frequently, but does not use any message more than once. In other words, if one sensor sends data very infrequently, the algorithm will have to wait and drop messages from all other sensors until it sees a new message from the low-frequency sensor to issue a join. The frequency of combined prediction is thus upper-bounded by the most infrequent sensor.
Such a wait can harm both end-to-end latency and accuracy due to the loss of high-frequency information.

\balance

\bibliographystyle{plain}
\bibliography{reference}
\newpage
\section*{Appendix}
\renewcommand{\thesubsection}{\Alph{subsection}}

\begin{figure*}[t]
    \captionsetup[subfigure]{justification=centering}
    \centering
    \begin{subfigure}{0.29\textwidth}
        \includegraphics[width=\columnwidth]{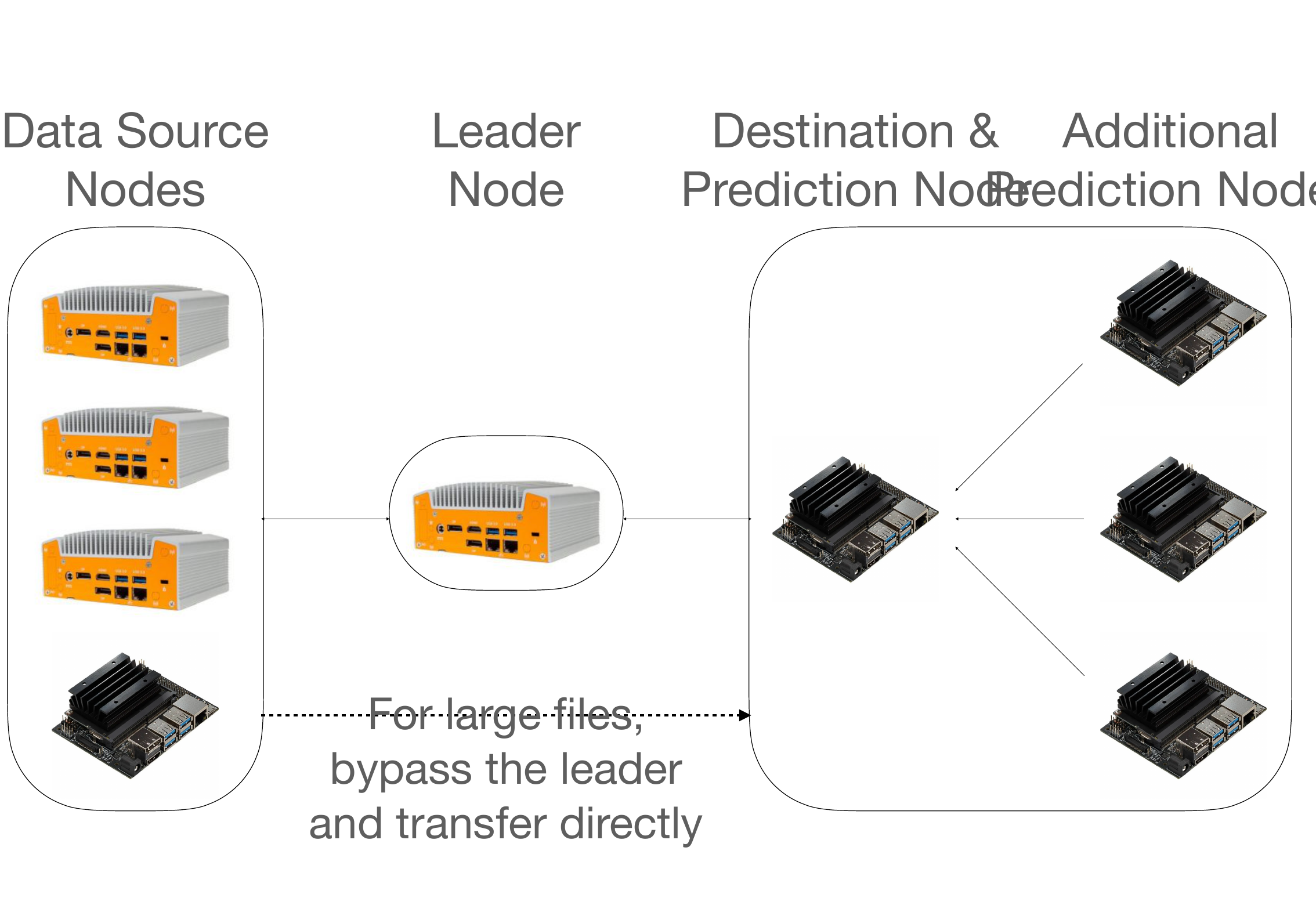}
        \caption{Topology 1\\ (Early fusion)}
        \label{fig:network-topology-1}
    \end{subfigure}
    \hfill
    \begin{subfigure}{0.38\textwidth}
        \includegraphics[width=\columnwidth]{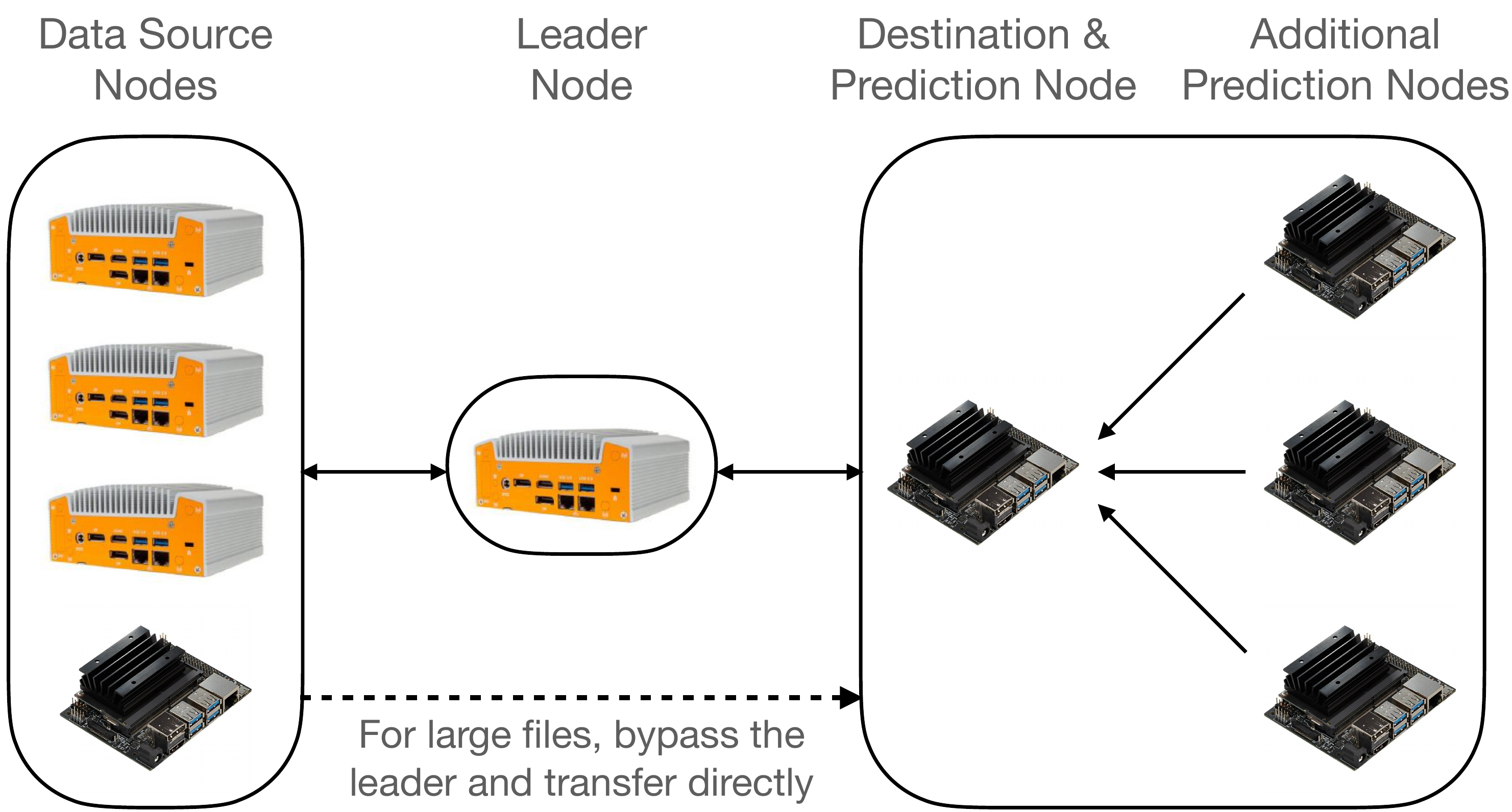}
        \caption{Topology 2\\ (Early fusion with parallelism)}
        \label{fig:network-topology-2}
    \end{subfigure}
    \hfill
    \begin{subfigure}{0.29\textwidth}
        \includegraphics[width=\columnwidth]{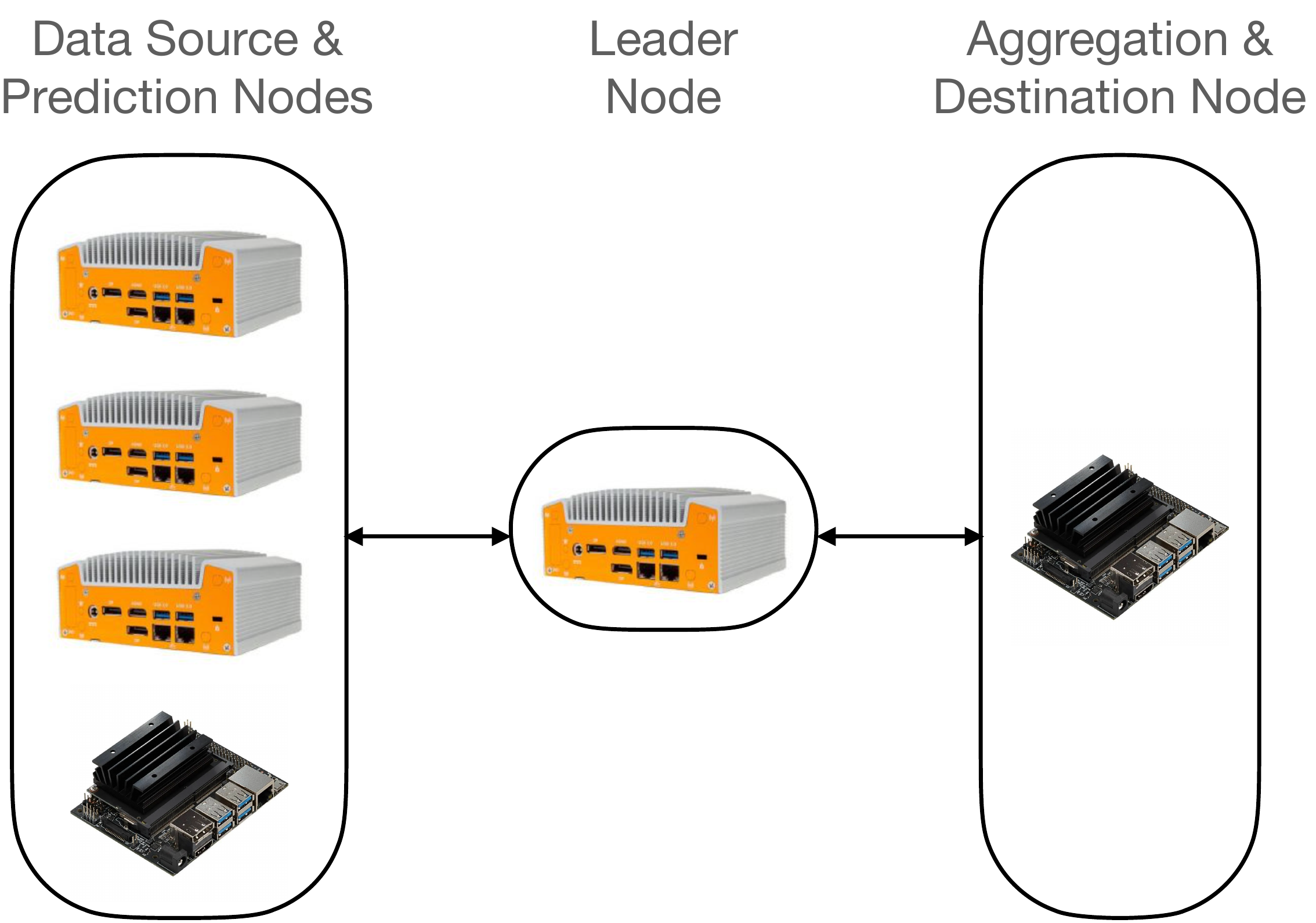}
        \caption{Topology 3\\ (Late fusion)}
        \label{fig:network-topology-3}
    \end{subfigure}
    \caption{Network topologies used in our experimental edge cluster. Topology 2 takes advantage of multiple prediction nodes consuming a shared message queue at the same time, and is only used in ``Parallel'' experiments.}
    \label{fig:network-topology}
\end{figure*}

\subsection{Model Decomposition}\label{sec:model-decomposition}
Since models are the unit of placement and computation in \sys, the goal of model decomposition is to increase opportunities for optimizing placement. The idea is to approximate a single model with an ensemble or mixture of smaller local models.
Obviously, not all models can be decomposed into smaller parts. However, many real-world models can be partitioned.

\vspace{0.5em} \noindent \textbf{Strategy 1. Ensemble Models}
Ensemble machine learning models are techniques that combine multiple models to improve the accuracy and robustness of predictions. 
Let's imagine that we have $p$ features and $n$ examples with an example matrix $X$ and a label vector $Y$.
Different subsets of these features are constructed on $m$ data sources on the network. Each source generates a partition of features $f_i$, i.e., $X[:, f_i]$ is the source-specific projection of training data. 
Stacking is an ensembling technique where multiple models are trained, and their predictions are used as inputs to a final model. The final model learns to weigh the predictions of each model and make a final prediction based on the weighted inputs. This helps capture the strengths of each individual model and produce a more accurate prediction.

We can train the following models. For each feature subset $f_i$, we train a model (from any model family) that uses only the subset of features to predict the label.
\[
g_i \leftarrow \textsf{train}(X[:, f_i], Y)
\]
After training each of these models over the $m$ subset, we train a stacking model that combines the prediction. This is a learned function of the outputs of each $g_i$ that predicts a single final label:
\[
h \leftarrow \textsf{train}([g_1,...,g_m], Y)
\]
Stacking models are well-studied in literature and are not new~\cite{sagi2018ensemble}. For multi-modal prediction tasks, prior work has found that such models do not sacrifice accuracy and sometimes actually improve accuracy~\cite{shaowang2023amir}.

\vspace{0.5em} \noindent \textbf{Strategy 2. Mixture of Experts Models}
Similarly, there are neural network architectures that can be trained end-to-end to take advantage of \sys. Mixture of Experts (MoE) is a deep learning architecture that combines multiple models or ``experts' to make predictions on a given task. The basic idea of the MoE architecture is to divide the input space into regions and assign an expert to each region. The gating network takes the input, decides which region it belongs to, and then selects the corresponding expert to make the prediction. The gating network then weights the output of each expert, and the final prediction is the weighted sum of the expert predictions. MoE architectures have been applied to a wide range of tasks, including language modeling, image classification, and speech recognition~\cite{eigen2013learning}.
After training, each expert can be placed independently once trained. 

\subsection{Evaluation Metrics}\label{sec:exp-metrics}
We borrow the following common metrics used in streaming systems~\cite{DBLP:conf/icde/KarimovRKSHM18} to measure the system performance of \sys.

\subsubsection{Types of Latency}\label{sec:exp-metrics-latency}

\noindent \textbf{Producer Sending Latency.} We define \textit{producer sending latency} to be the interval between the time the producer begins sending an example to the leader node and the time the producer finishes sending the same example to the leader node.

\noindent  \textbf{Consumer Receiving Latency.} We define \textit{consumer receiving latency} to be the interval between the time the producer finishes sending an example to the leader node and the time the consumer finishes receiving the same example from the leader node. That means, if there is any queuing backlog at the leader node, it is counted as part of \textit{consumer receiving latency}.

\noindent \textbf{Total Communication Latency.} We define \textit{total communication latency} to be the sum of \textit{producer sending latency} and \textit{consumer receiving latency}. It means the interval between the time the producer begins sending an example to the leader node and the time the consumer finishes receiving the same example from the leader node.

\noindent \textbf{Reaction Time.} We define \textit{reaction time} to be the interval between the time the producer begins sending the latest example involved in a join and the time at which the joined data tuple arrives at the consumer node. It measures how timely the system reacts to the newest information available.

\noindent  \textbf{Processing Latency.} We define \textit{processing latency} to be the interval between the time a prediction node starts processing an example (or a joined set of examples) and the time it finishes processing the same example. The processing latency is used to measure the actual computation time of an example (or a joined set of examples).

\noindent  \textbf{End-to-end Latency.} We measure the interval between the time an example is collected by \sys and the time the last prediction node finishes processing the same example as \textit{end-to-end latency}. The end-to-end latency includes but not limited to total communication latency and processing latency. 

\noindent \textbf{Queueing Time.} We define \textit{queueing time} to be the interval between the time a node finishes its local inference of an example and the time such local prediction is joined with other local outputs. It measures how long it has to wait for other local predictions to be joined together.

 \noindent \textbf{Total Working Duration.} We define \textit{total working duration} to be the interval between the time the producer begins sending the first example of a task to the leader node and the time the last prediction node finishes processing the last example of the task. This is a task-level measurement rather than a per-example measurement, and it includes both communication time and computation time (if any).

\subsubsection{Backlog}\label{sec:exp-metrics-backlog}
We define the \textit{end-to-end latency} of the last example of a task as \textit{backlog}.
Backlog is an important metric because all kinds of delays can easily accumulate, which causes outdated predictions for later examples in a real-time inference scenario.
The lower bound of the backlog is near-zero, when there is no delay along the path from the data source to prediction nodes.
Ideally, such lower bound is achievable if data arrive slower than the rate our computational power can serve, or we might have to skip some data points to keep the predictions in time.

\subsection{Benefits of Late Fusion Models}\label{sec:appendix-late-fusion}
There has been recent work discussing the latency vs. accuracy tradeoff between early fusion and late fusion models~\cite{wayformer}.
Early fusion models, while potentially capturing more cross-modal correlations, tend to require more communication as they combine raw data from various sources.
In contrast, late fusion models reduce communication by combining locally inferred results rather than raw data.
Late fusion models also naturally support parallelism since most inference is performed locally.
For latency-sensitive tasks, the communication efficiencies offered by late fusion models are extremely valuable.
We conduct experiments to demonstrate the benefits of late fusion models in \sys.

\subsubsection{Network Topology Setup}
Figure~\ref{fig:network-topology} shows the network topologies of our experiment setup.
Network topology 1 in Figure~\ref{fig:network-topology-1} uses only one prediction node, which is supported by both \sys and PyTorch.
Network topology 2 in Figure~\ref{fig:network-topology-2} has 3 additional prediction nodes, and all these 4 prediction nodes consume a shared queue at the same time in parallel experiments thanks to \sys. This cannot be done in PyTorch due to the lack of a shared queue.
Network topology 3 in Figure~\ref{fig:network-topology-3} takes advantage of model decomposition described in \S\ref{sec:model-decomposition} and uses local data source nodes as local prediction nodes, too. The node that was making prediction in topology 1 and 2 now only has to gather local predictions and take a majority vote.
All topologies have 4 data source nodes, from which we use 3 NUCs (in yellow) and 1 Jetson Nano (in black) to reflect the heterogeneity of real-world edge devices. The other NUC is set up as the leader node, or the ``master'' node in terms of PyTorch distributed. The rest of Jetson Nanos are prediction nodes where the actual computation is done, and one of them is designated as the destination node where final results should be sent.

\subsubsection{Late Fusion Models Reduce Backlog}
\begin{figure}[htbp]
    \centering
    \includegraphics[width=0.9\columnwidth]{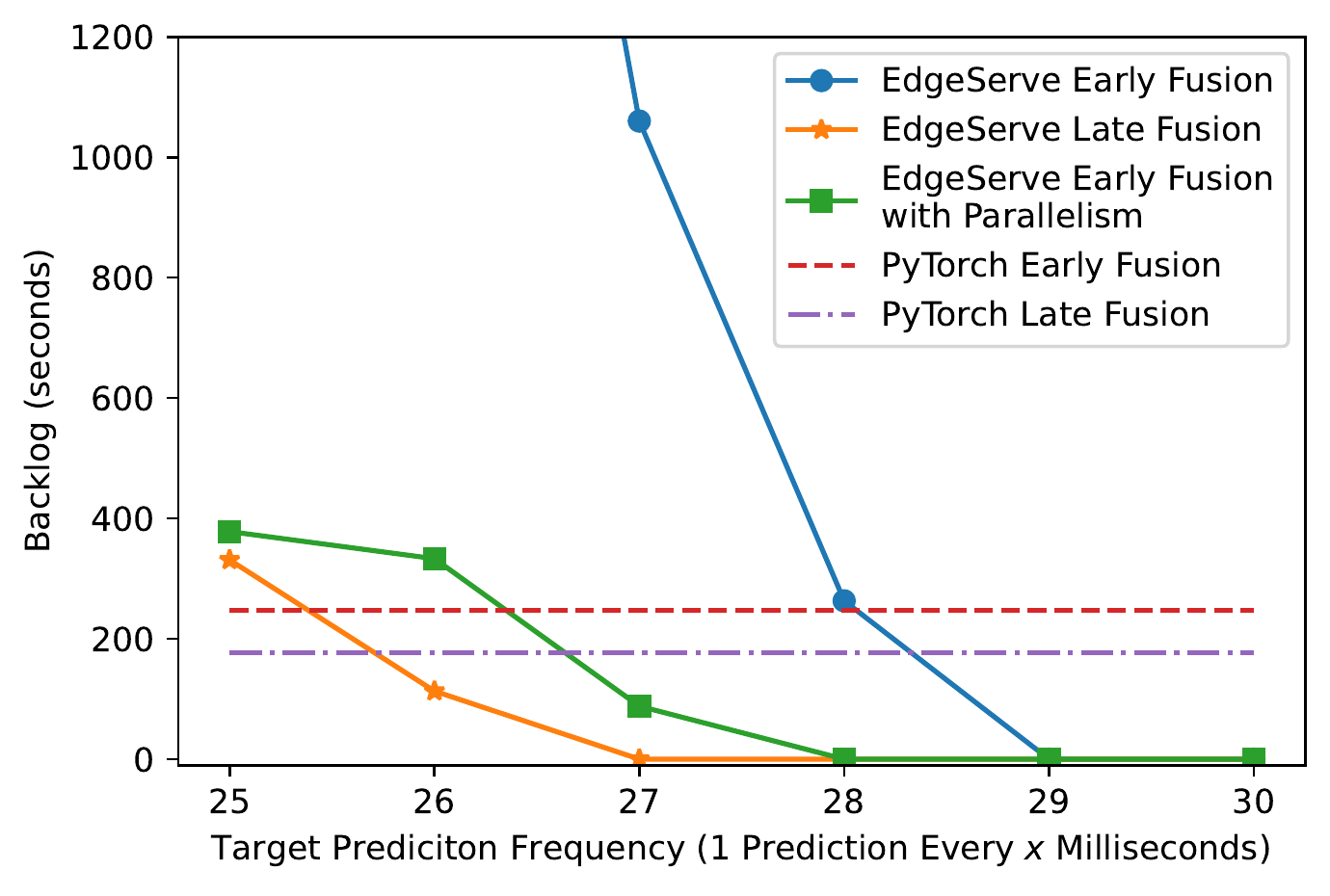}
    \caption{Measure of backlog in the activity recognition task for all network topologies.}
    \label{fig:exp-opportunity-latency-full}
\end{figure}

Figure~\ref{fig:exp-opportunity-latency-full} is an extension to Figure~\ref{fig:exp-opportunity-latency}, showing the backlog for all network topologies.
For both \sys and PyTorch, we see a lower backlog for late fusion models (Topology 3) because we are able to make the most of local data source nodes and save communication costs.
\sys early fusion with parallelism (Topology 2) also helps reduce the backlog over early fusion (Topology 1), when the model is not able to catch up with incoming data rate.

\subsubsection{Late Fusion Models Are More Tolerant To Delays}
\begin{table}[]
\centering
\small
\begin{tabular}{l|l|l}
Real-time accuracy (F-1 score)   & No delay  & 25ms delay \\ \hline
\sys Early Fusion           & 0.90   & 0.55     \\
\sys Early Fusion w/ Parallelism              & 0.90   & 0.55     \\
\sys Late Fusion         & 0.91   & 0.85
\end{tabular}
\caption{Real-time accuracy measured in F-1 score when one of the four data streams has a constant delay. Late fusion models are more accurate even if there is a delay from one data source.}
\label{tab:exp-opportunity-delayed-accuracy}
\end{table}
We next evaluate the fail-soft benefits of \sys, by measuring real-time accuracy of predictions when one of the four data streams have a constant 25ms delay. Table~\ref{tab:exp-opportunity-delayed-accuracy} shows that, while early fusion models (with or without parallelism) suffer from considerable degradation in accuracy due to the delay, late fusion models achieves a much higher accuracy even if there is a constant delay. This is because local predictions from other data streams are unaffected and the ``unpopular vote'' -- local prediction of the delayed data stream is likely dropped by the ensemble method.

\begin{figure}[t]
    \centering
    \includegraphics[width=0.9\columnwidth]{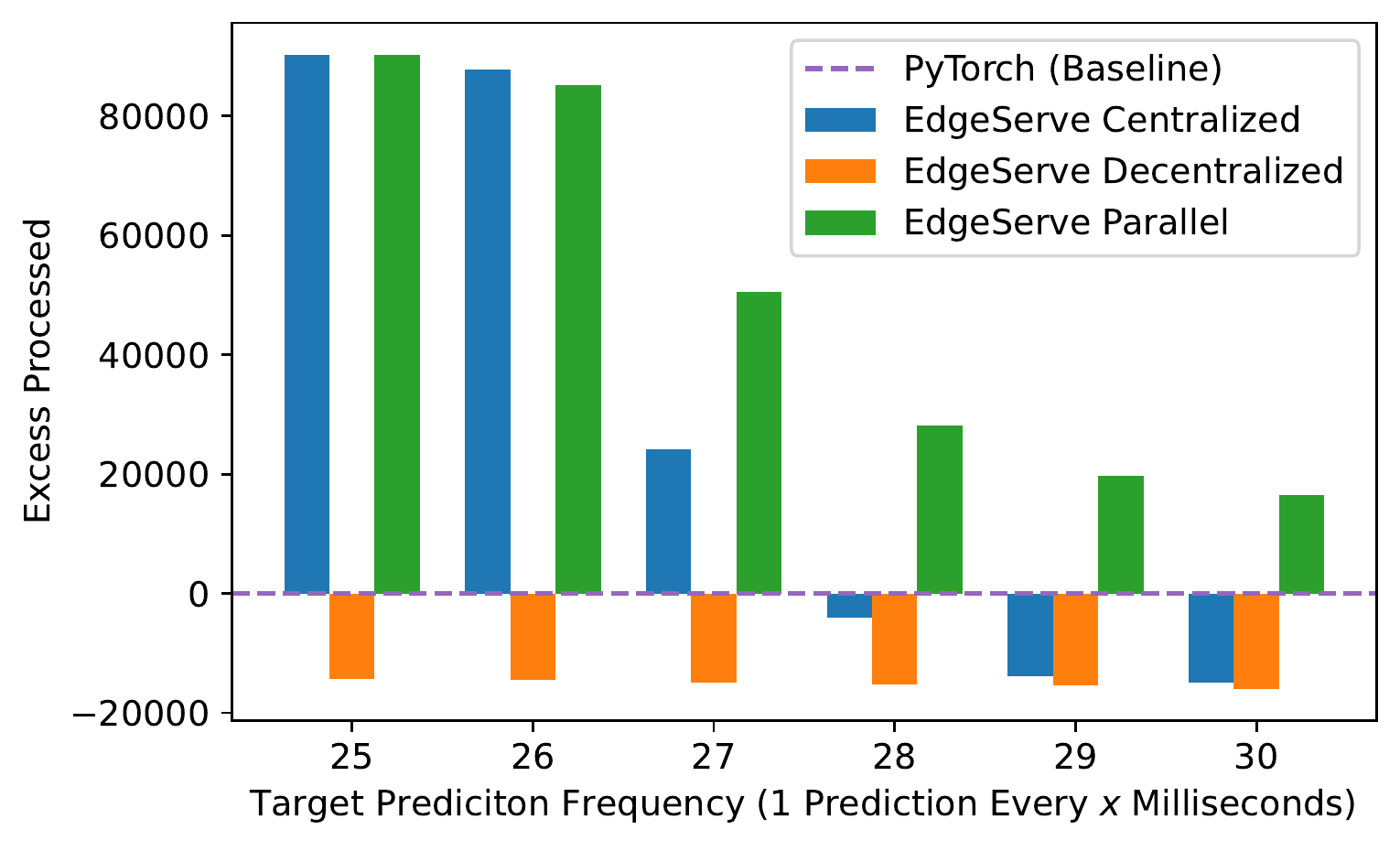}
    \caption{Number of excess examples processed for different strategies and target prediction frequencies.}
    \label{fig:exp-opportunity-examples}
\end{figure}
\subsubsection{Late Fusion Models Reduce Excess Work.}
Now, we look at the number of excess examples that are processed to better understand how \sys automatically downsamples the incoming data stream in response to the target prediction frequency.
As can be seen from Figure~\ref{fig:exp-opportunity-examples}, PyTorch (either early or late fusion), as a baseline, is marked as zero on the y-axis because it always processes a fixed number of examples equal to the input size.
In early fusion (with or without parallelism) settings, \sys is very sensitive to target prediction frequency because it can downsample incoming data stream when such target is relaxed. Therefore, we see a rapidly decreasing excess work from left to right as the target prediction frequency becomes less frequent.
On the other hand, in a late fusion setting, \sys is not as sensitive to such change in target prediction frequency for the same reason why \sys late fusion maintains a high real-time accuracy discussed in \S\ref{sec:exp-opportunity}. After faster NUCs finish local predictions, the ensemble model simply skips further local predictions made by the Jetson Nano as they fall outside the acceptable time skew range.
As a result, late fusion models only process a small number of examples, even when the target prediction frequency is high enough.

\subsection{Comparison with Federated learning.}
Our work on decentralized prediction might seem similar to federated learning~\cite{DBLP:journals/corr/KonecnyMR15, MLSYS2019_bd686fd6}, but there are several key differences. First, our goal is not to collaboratively train a shared model, but to make combined predictions based on multiple streams of data. Second, we optimize for millisecond-level end-to-end timeliness from the point of data collection to the point where prediction is delivered. Federated learning tasks usually assume a much longer end-to-end latency, and they have other optimization goals, such as communication cost. Third, we have to take care of time-synchronization between data streams, while federated learning systems usually treat those data as the same batch.

\end{document}